# Flux Free Single Crystal Growth and Detailed Physical Property Characterization of $Bi_{1-x}Sb_x$ (x = 0.05, 0.1 and 0.15) Topological Insulator

Rabia Sultana[1,2], Ganesh Gurjar[3], Bhasker Gahtori[1,2], Satyabrata Patnaik[3] and V.P.S. Awana[1,2*]

[1] National Physical Laboratory (CSIR), Dr. K. S. Krishnan Road, New Delhi-110012, India

[2] Academy of Scientific and Innovative Research (AcSIR), Ghaziabad-201002, India

[3]School of Physical Sciences, Jawaharlal Nehru University, New Delhi-110067, India

**Abstract**

Here, we report the crystal growth, physical and transport properties of $Bi_{1-x}Sb_x$ (x = 0.05, 0.1 and 0.15) topological insulator. Single crystals of $Bi_{1-x}Sb_x$ (x = 0.05, 0.1 and 0.15) were grown by melting bismuth and antimony together using the facile self flux method. The XRD measurements displayed highly indexed 00l lines and confirmed the crystalline nature as well as the rhombohedral structure of the $Bi_{1-x}Sb_x$ (x = 0.05, 0.1 and 0.15) crystals. Raman spectroscopy measurements for $Bi_{1-x}Sb_x$ system revealed four peaks within the spectral range of 10 to 250 $cm^{-1}$ namely $A_{1g}$ and $E_g$ modes corresponding to Bi-Bi and Sb-Sb vibrations. Scanning electron microscopy (SEM) and energy dispersive X-ray analysis (EDAX) measurements showed the layered surface morphology and near stoichiometric chemical composition of $Bi_{1-x}Sb_x$ (x = 0.05, 0.1 and 0.15) crystals. Furthermore, EDAX mapping confirmed the homogeneous distribution of Bi and Sb elements. Temperature dependent electrical resistivity curves with and without applied magnetic field exhibited a metallic behaviour and linear non-saturating magneto-resistance (MR) respectively for all the antimony (Sb) concentrations of x = 0.05, 0.1 and 0.15. The lowest Sb concentration sample with x = 0.05 ($Bi_{0.95}Sb_{0.05}$) exhibited the highest MR value of about 1400%, followed by x = 0.1 and 0.15 samples ($Bi_{0.9}Sb_{0.1}$ and $Bi_{0.85}Sb_{0.05}$) with MR values reaching up to 500% and 110% respectively at 2K and 6Tesla applied field. Also, a coexistence of negative MR and WAL/WL behaviour is observed at lower magnetic fields (below ± 0.2Tesla) in $Bi_{0.9}Sb_{0.1}$ and $Bi_{0.85}Sb_{0.05}$ system. To further elaborate the transport properties of $Bi_{1-x}Sb_x$ (x = 0.05, 0.1 and 0.15), the magneto-conductivity (MC) is fitted to the HLN (Hikami Larkin Nagaoka) equation and it is found that the charge conduction mechanism is mainly dominated by WAL (weak anti-localization) along with a small contribution from WL (weak localization) effect. Summarily, the short letter discusses the synthesis, interesting transport and magneto-transport properties of $Bi_{1-x}Sb_x$ (x = 0.05, 0.1 and 0.15), which could be useful in understanding the fascinating properties of topological insulators and their technological applications.

**[*]Corresponding Author**

Dr. V. P. S. Awana: E-mail: awana@nplindia.org
Ph. +91-11-45609357, Fax-+91-11-45609310
Homepage: awanavps.webs.com

## Introduction

The advancement in the field of condensed matter physics is a result of the discovery of novel materials with exotic properties. Recently, the discovery of various Dirac materials viz., Graphene, three dimensional (3D) Topological Insulators (TIs) and Topological (Dirac/Weyl) semi-metals has attracted immense interest concerning both fundamental research and device applications [1-3]. In particular, 3D TIs, one of the newest wonders in condensed matter physics community has attracted significant attention due to their relatively large bulk band gaps and topologically protected surface states along with a single Dirac cone at the Γ point of the Brillouin zone [1, 4-13]. Further, strong spin–orbit coupling (SOC) along with time reversal symmetry (TRS) protected surface states makes these TIs as potential candidates for understanding novel quantum phenomena as well designing of futuristic electronic devices [1, 4-13]. To date, we have witnessed rapid research developments both on the side of experimental as well as in the theoretical understanding of these novel materials i.e., TIs. However, the realisation of 3D TIs along with fascinating properties is still a challenging topic. Indeed, some of the well - known three dimensional (3D) TIs discovered so far include $Bi_{1-x}Sb_x$, $Bi_2Se_3$, $Bi_2Te_3$, $Sb_2Te_3$ and so on [1, 4, 5, 10-13]. Interestingly, all the confirmed 3D TIs are chalcogenides i.e., possessing chalcogen atoms (Se and Te), whereas, $Bi_{1-x}Sb_x$ is an alloy. In this work, we report our investigations in one such TI material, namely $Bi_{1-x}Sb_x$.

Following the theoretical discovery in early 1980's, the two dimensional (2D) TI exhibiting quantum spin Hall (QHE) effect was experimentally observed in HgTe/ CdTe quantum well [8,9,14,15]. This further led to the experimental discovery of the first 3D TI i.e., $Bi_{1-x}Sb_x$ in 2008 by Hsieh et.al [10]. Its unusual surface bands were experimentally mapped using the angle resolved photoemission spectroscopy (ARPES) and scanning tunnelling spectroscopy (STS) [10-13]. $Bi_{1-x}Sb_x$ is a semiconducting alloy of bismuth (Bi) and antimony (Sb) and is widely known for its thermoelectric properties [16]. Pure Bi is a semi-metal possessing strong SOC but, when substituted with Sb, the critical energies of the band structure changes. As reported, several physical factors (alloy composition, temperature, external pressure and magnetic field) govern the band structure of $Bi_{1-x}Sb_x$ system [14-19]. Also, several reports have discussed theoretically about the electronic structure as well as the surface states of $Bi_{1-x}Sb_x$ based on the first principles calculations [20, 21]. Accordingly, $Bi_{1-x}Sb_x$ material exhibits band inversion at an odd number of time reversal invariant momentum (TRIM) and has an opening of a bulk band gap in the Sb concentration range of 0.09 to 0.23 [5,16,17]. Additionally, $Bi_{1-x}Sb_x$ exhibits different states depending upon the Sb concentration viz., TI state (x = 0.03 - 0.22), Dirac/Weyl semi-metal (DSM/WSM) below x =0.03 [3, 10, 12, 21-24]. It is interesting to note that when the TRS is broken, DSM changes to WSM and exhibit various exciting transport properties such as large carrier mobility, low carrier density, anomalous Hall effect (AHE), giant linear magneto-resistance (MR), and negative longitudinal MR (LMR) [2,3,24-27]. As reported, the negative LMR is a consequence of the Adler-Bell –Jackiw Chiral anomaly and occurs when the magnetic field and electric field are parallel to each other (B||E) [2, 3, 24-34]. So far, the chiaral anomaly induced negative LMR feature has been observed in 3D TIs, DSMs and WSMs ($Bi_{1-x}Sb_x$, $Cd_3As_2$, $Na_3Bi$, $ZrTe_5$, TaAs and TaP) [26-34]. Further, Shin et al., have observed the violation of Ohm's law in $Bi_{1-x}Sb_x$ (x=0.04) sample for E||B in the Weyl metal state [35]. More recently, Su et al. theoretically investigated two WSM phases in $Bi_{1-x}Sb_x$ (x=0.5 and 0.83) samples [36]. Owing to its rapid progress, it is thus predicted that more novel properties will be discovered in the coming future in this ($Bi_{1-x}Sb_x$) well studied systems. Thus, it is very crucial to understand the fundamental physical properties of $Bi_{1-x}Sb_x$ TI for practical use. In the present paper, we discuss the growth of quality $Bi_{1-x}Sb_x$ (x = 0.05, 0.1 and 0.15) single crystals and study the physical and transport properties of the same.

**Experimental Details**

$Bi_{1-x}Sb_x$ (x = 0.05, 0.1 and 0.15) single crystals were grown using the self flux method via the conventional solid state reaction route. Stoichiometric mixtures (~1 gram) of high purity (99.999%) Bi and Sb were weighed accurately and grounded thoroughly inside an Argon filled glove box to obtain a homogeneous mixture. The well mixed powder after pelletization was sealed in an evacuated quartz tube ($10^{-3}$Torr) and kept horizontally inside an automated programmable tube furnace. The quartz tube was heated to 650˚C with a rate of 42˚C/hour, hold for 8 hours and then slowly cooled (3˚C/hour) to 250˚C with a hold time of 95hours. Further, the sample was cooled from 250˚C to 245˚C (3˚C/hour), kept there for 73 hours and then finally allowed to cool down to room temperature (120˚C/hour). The detailed heat treatment is shown in the schematic diagram [Figure 1]. The as grown crystals (~1 cm) were mechanically cleaved, resulting into a silvery reflective surface, which was prepared for further characterizations [Figure2].

The primary structural characterization of the synthesized samples were carried out by employing Rigaku Miniflex II, Desktop X-ray Diffractometer (XRD) with Cu-Kα radiation (λ=1.5418 Å). The Rietveld refinements of the powder XRD raw data were performed using the Full Prof Suite Toolbar software. Visualization for Electronic and Structural analysis (VESTA) software was used to determine the formation and analysis of the unit cell of different composition of $Bi_{1-x}Sb_x$ samples. The surface characteristics and constituent elements mapping of the synthesized $Bi_{1-x}Sb_x$ (x = 0.05, 0.1 and 0.15) samples were analysed by using ZEISS-EVO MA-10 scanning electron microscopy (SEM) images coupled with energy dispersive X-ray analysis (EDAX). Raman spectra of as grown $Bi_{1-x}Sb_x$ (x = 0.05, 0.1 and 0.15) single crystals were recorded at room temperature using the Renishaw Raman Spectrometer. The electrical and magnetic measurements were performed using the Cryogenics System under applied magnetic fields of up to 6Tesla and temperature down to 2K. Additionally, magneto-conductivity (MC) of the synthesized samples was fitted to the HLN (Hikami Larkin Nagaoka) model with the help of MATLAB R2015a software.

**Results and Discussion**

To analyze the crystalline nature, XRD measurements were conducted on freshly cleaved $Bi_{1-x}Sb_x$ (x = 0.05, 0.1 and 0.15) single crystals. Figure 3 (a, b ,c) displays the XRD patterns carried out on the freshly cleaved silvery surface of three $Bi_{1-x}Sb_x$ crystals with increasing Sb concentration viz., x = 0.05, 0.1 and 0.15 in the angular range of $2\theta_{min} = 5^\circ$ and $2\theta_{max} = 80^\circ$. All the three samples of as synthesised $Bi_{1-x}Sb_x$ ranging from x= 0.05 to 0.15 exhibited sharp (00l) reflections which are well indexed along the c-axis, indicating the good single crystalline nature. However, an additional peak is observed to appear at $2\theta \approx 49.5^\circ$ along with the characteristic peaks of (00l) plane for all the three compositions ($Bi_{0.95}Sb_{0.05}$, $Bi_{0.9}Sb_{0.1}$ and $Bi_{0.85}Sb_{0.15}$). For the $Bi_{0.9}Sb_{0.1}$ sample [Fig. 3(b)] the additional peak observed exhibits much lower intensity as compared to the other two composition [Figure 3(a,c)]. The additional peak is identified as (2 0 2) and marked by asterisk as depicted in Fig. 3(a, b, c). The existence of (2 0 2) peak might be due to the misaligned plane of $Bi_{1-x}Sb_x$ (x = 0.05, 0.1 and 0.15) samples.

In order to confirm the phase purity and crystal structure of the as grown single crystals, powder XRD (PXRD) patterns on finely crushed crystals of $Bi_{1-x}Sb_x$ (x = 0.05, 0.1 and 0.15) were collected at room temperature. Figure 4(a, b, c) depicts the Rietveld refinement of as synthesized $Bi_{1-x}Sb_x$ (x = 0.05, 0.1 and 0.15) samples. The fitted patterns (red solid lines) match well with the experimentally observed patterns (black solid lines).

Consequently, all the $Bi_{1-x}Sb_x$ (x = 0.05, 0.1 and 0.15) samples exhibited c-axis oriented rhombohedral structure of the space group, R-3m. The refined lattice parameters, atom parameters as well as the Chi square values are displayed in Table 1. As seen from Table 1, with increase in Sb concentration from x = 0.05 to 0.15, the value of the lattice parameter c is observed to decrease from 11.81(2) Å to 11.77(5) Å.

Figure 5 (a, b, c) represents the unit cell structure of as synthesized $Bi_{1-x}Sb_x$ (x = 0.05, 0.1 and 0.15) single crystals formed using the VESTA software. Although, $Bi_{1-x}Sb_x$ sample exhibits rhombohedral crystal structure similar to the earlier reported TIs ($Bi_2Se_3$, $Bi_2Te_3$, $Sb_2Te_3$) [37, 42, 43], the arrangement of the atoms in the unit cell structure is different. Here, all the unit cells contain three bi-layers of Bi and Sb stacked one over the other, as depicted in Fig.5 (a, b, c).

To determine the surface morphology as well as the compositional analysis of as grown $Bi_{1-x}Sb_x$ (x = 0.05, 0.1 and 0.15) single crystals scanning electron microscopy (SEM) and energy dispersive X-ray analysis (EDAX) mapping was done. Figure 6 (a, b, c) depicts the SEM and EDAX mapping images taken on freshly cleaved $Bi_{1-x}Sb_x$ (x = 0.05, 0.1 and 0.15) single crystals. The morphology of all the $Bi_{1-x}Sb_x$ (x = 0.05, 0.1 and 0.15) samples showed the layered growth, which appeared as smooth plate like laminar structure [Fig. 6 (a, b, c) (i)] Thus, the SEM images confirmed the layered nature of TIs. Further, the elemental analysis was carried out using EDAX on small portion of SEM image of $Bi_{1-x}Sb_x$ (x = 0.05, 0.1 and 0.15) samples. The EDAX analysis displays the presence of constituent elements (Bi and Sb only) in nearly stoichiometric ratio and confirmed that the as synthesized $Bi_{1-x}Sb_x$ crystals are of pure form i.e., free from foreign contamination. Also, the quantitative weight percentage values of the atomic constituents (Bi and Sb) for all the three samples are displayed in the Fig. 6(a, b, c) ii. The elemental mapping of all the $Bi_{1-x}Sb_x$ (x = 0.05, 0.1 and 0.15) crystals showed that the constituent elements i.e. Bi and Sb are homogeneously distributed according to their stoichiometric ratio i.e., composition close to the $Bi_{1-x}Sb_x$ (x = 0.05, 0.1 and 0.15) samples [Fig. 6(a, b, c) iii and Fig. 6(a, b, c) iv].

To observe the substitution effect of Bi by Sb atoms in $Bi_{1-x}Sb_x$ system and to confirm the composition dependence of the vibrational modes of as grown $Bi_{1-x}Sb_x$ (x = 0.05, 0.1 and 0.15) single crystals, Raman spectroscopic measurements were performed using a Renishaw Raman Spectrometer. The Raman spectra were taken on mechanically cleaved silvery shiny surface of as obtained $Bi_{1-x}Sb_x$ single crystals. Room temperature Raman spectra of all the compositions of $Bi_{1-x}Sb_x$ (x = 0.05, 0.1 and 0.15) system are recorded using a laser excitation wavelength of 514 nm along with a spectral resolution of 0.5cm$^{-1}$. In order to avoid the sample surface burning and damage, the laser power was maintained below 5mW. Figure 7 (a, b, c) illustrates the Raman spectra of $Bi_{1-x}Sb_x$ (x = 0.05, 0.1 and 0.15) single crystals. Clearly, four distinct phonon modes are observed in the measured spectral range of 10-250 cm$^{-1}$. The corresponding values of these four modes are displayed in Table 2. Interestingly, the four vibrational modes observed for $Bi_{1-x}Sb_x$ system differs from the pure ($Bi_2Te_3$, $Bi_2Se_3$ and $Sb_2Te_3$) TI system (reported earlier by our own group), which exhibited only three distinct Raman peaks [37]. As reported, pure Bi and Sb exhibits two types of Raman active modes i.e., $A_{1g}$ (upper frequency mode) and $E_g$ (lower frequency mode [38 - 41].The $A_{1g}$ mode is singly degenerate, whereas the $E_g$ mode is doubly degenerate. In the present $Bi_{1-x}Sb_x$ (x = 0.05, 0.1 and 0.15) system, the characteristic peaks observed at ~72 - 75cm$^{-1}$ and ~97 - 100 cm$^{-1}$ corresponds to the Bi-Bi vibrations of the $E_g$ and $A_{1g}$ modes respectively. On the other hand, the other two peaks observed at ~118 - 120cm$^{-1}$ and ~138 – 141 cm$^{-1}$ represent the Sb - Sb vibrations of the $E_g$ and $A_{1g}$ modes [Fig. 7 (a, b, c)]. It may be noted that as the Sb concentration increases from x = 0.05 to 0.15 in $Bi_{1-x}Sb_x$ system, the values of $E_g$ and $A_{1g}$ mode frequencies corresponding to Bi-Bi and Sb-Sb vibrations tends to increase (see Table 2). Accordingly, the $Bi_{1-x}Sb_x$ system exhibits a total of four $E_g$ and $A_{1g}$ peaks corresponding to Bi-Bi and Sb-Sb vibrations and the values of these mode frequencies are in agreement to the earlier reported literature [38 - 41].

Figure 8 (a, b, c) displays the temperature dependent resistivity plots for $Bi_{1-x}Sb_x$ (x = 0.05, 0.1 and 0.15) single crystals under different applied perpendicular magnetic fields. All the three graphs of Fig. 8 (a, b, c) representing $Bi_{0.95}Sb_{0.05}$, $Bi_{0.9}Sb_{0.1}$ and $Bi_{0.85}Sb_{0.15}$ exhibits metallic nature both in the absence as well as in the presence of applied magnetic field i.e., resistivity increases with increase in temperature. The inset of Fig. 8 (a, b, c) shows the Fermi liquid behaviour of $Bi_{1-x}Sb_x$ (x = 0.05, 0.1 and 0.15) samples at 0Tesla. The Fermi liquid fitting was done using the formula $\rho = \rho_0 + AT^2$, where $\rho$, $\rho_0$, A and T represents the resistivity, residual resistivity, constant and the temperature respectively. Residual resistivity ($\rho_0$) is defined as the resistivity at zero temperature (0K) and is obtained by extrapolation. Both the residual resistivity ($\rho_0$) as well as the constant (A) value is displayed in Table 3. All the resistivity curves [Fig.8 (a, b, c)] follows the Fermi liquid behaviour in a temperature range from 5K to about 50K.

Figure 9 (a, b, c) represents the percentage change of magneto- resistance (MR) as a function of applied magnetic field (0 to 6Tesla) at different temperatures (2 to 200K) of as grown $Bi_{1-x}Sb_x$ (x = 0.05, 0.1 and 0.15) single crystals. Here, the magnetic field was applied perpendicular to the current flow. In order to obtain symmetric MR values, we have taken the average of the resistivity raw data over positive and negative field directions as $\rho_{sym} = [\rho_{raw}(+H) + \rho_{raw}(-H)] / 2$ [42]. Thereafter, the MR (%) for all the studied $Bi_{1-x}Sb_x$ samples were calculated using the formula MR= $[\rho_{sym}(H) - \rho_{sym}(0)]/\rho_{sym}(0)$ where, $\rho_{sym}(H)$ and $\rho_{sym}(0)$ specifies the symmetric resistivity under applied and zero magnetic field respectively. The data used for Fig. 9(a) is the as acquired data. On the contrary, for Fig. 9(b-c) we did not observe any oscillations rather a little bit noisy MR plots were obtained from the as acquired data. Further, to compensate the noise, the MR data thus obtained from the as acquired data was slightly smoothened. The calculated MR (%) values of all the three $Bi_{1-x}Sb_x$ samples are displayed in Table 4.

Figure 9(a) reveals the MR (%) behaviour of $Bi_{0.95}Sb_{0.05}$ crystal. Apparently, at lower temperatures (2, 5 and 10K) the MR curves appears to overlap on one another and the shape of the plots are nearly V type i.e., weak anti localization (WAL) behaviour is observed near zero magnetic fields (H < 0.5Tesla). Further, as the temperature increases from 10 to 200K the MR curves gradually broadens i.e., changes from nearly V shape to U shape at magnetic fields (H < 0.5Tesla). Accordingly, $Bi_{0.95}Sb_{0.05}$ single crystal is observed to exhibit the largest, high field non-saturating positive MR (%) value reaching up to ~1400% at 2K and further decreases to ~500% as the temperature is increased to 200K [Fig. 9(a)]. This high value of MR (1400%) at the lowest measured temperature (2K) obtained for $Bi_{0.95}Sb_{0.05}$ single crystal seems interesting from both basic research and application point of view.

On the other hand, $Bi_{0.9}Sb_{0.1}$ sample exhibited much less value as compared to the $Bi_{0.95}Sb_{0.05}$ sample i.e., the MR (%) decreases from~ 440% at 2K to ~130% at 200K respectively [Fig. 9b(i)]. Similar to $Bi_{0.95}Sb_{0.05}$ sample, the MR plots of $Bi_{0.9}Sb_{0.1}$ sample also seems to overlap at lower temperatures (2, 5 and 10K) which gradually tend to broaden at higher temperatures (50,100 and 200K). Indeed, an interesting behaviour is observed at lower temperatures (2 to 10K) and magnetic field (H < ± 0.2Tesla) see fig. 9b (ii). At 10K, the MR plot reveals a positive MR with a sharp WAL behaviour [Fig. 9b (ii)]. But, at further lower temperatures (2 and 5K), a negative MR with weak localization (WL) behaviour is observed. Above ± 0.08Tesla (for 2K) and ± 0.18Tesla (for 5K), the MR curve takes an upturn with positive MR value and the WL behaviour disappears. Clearly, the coexistence of WL behaviour and negative MR at lower temperatures and magnetic field is really amazing. Generally, materials with strong SOC exhibits positive MR value with WAL behaviour at lower temperatures and applied perpendicular magnetic fields [37, 43, 44]. In our present case, the possible reason for the coexistence of negative MR as well as WL behaviour might be due to the slight misalignment of the direction of the applied magnetic field and the transport current. The actual underlying mechanism behind the coexistence of negative MR and WL behaviour needs further studies.

Additionally, the $Bi_{0.85}Sb_{0.15}$ sample exhibited unusual MR behaviour with values ranging from ~60% at 2K to ~52% at 50K respectively [Fig. 9c (i)]. Also, this ($Bi_{0.85}Sb_{0.15}$) sample shows some peculiar behaviour above 5Tesla for 2, 5, 10 and 50K respectively i.e., a knee-like kink feature appears above an applied magnetic field of 5Tesla. Further, a steep increase in the MR behaviour is observed above 5Tesla. It is well known that at lower applied magnetic fields the MR exhibits a quadratic response, whereas at higher fields the MR is dominated by linear field dependence along with a small quadratic term [MR= MR= aH+ b($H^2$)] [44]. To understand the mechanism behind the unusual MR of $Bi_{0.85}Sb_{0.15}$ sample, we fitted the MR data using the quadratic equation MR= aH+ b($H^2$) as shown in fig. 9c (ii). Apparently, the MR data at 2, 5 and 10K fits well to the above quadratic relation below 5Tesla, which suggests that the MR originated from contributions of both the quadratic and linear terms. However, the asymptotic increase of MR above 5Tesla needs further investigation in terms of higher field measurements.

Moreover, a coexistence of WAL/WL along with negative MR was observed at lower applied magnetic fields (H < ± 0.5Tesla) and temperatures below 50K as illustrated in fig. 9c (iii). The MR curve at 2K revealed a narrow dip (extremely sharp positive v-type cusp) near zero applied magnetic field which can be attributed to the WAL effect. Initially, the MR curve at 2K is observed to be increasing with positive values up to ± 0.05Tesla, and then starts to decrease at even higher fields (up to ± 0.2Tesla) with negative values. Further, above ± 0.2Tesla the MR curve takes an upturn and exhibits positive values. With the increment in temperature, the localization behaviour of the $Bi_{0.85}Sb_{0.15}$ sample is significantly changed. As the temperature increases from 2K to 5K, the MR data is observed to broaden slightly i.e., low field v-type cusp is not as sharp as that observed for 2K, suggesting weakened WAL effect. The MR observed at 5K exhibits positive values up to ± 0.04Tesla, which further decreases with increase in magnetic field up to ± 0.16Tesla and then finally takes an upturn in a similar fashion to that observed for 2K. Above ± 0.2Tesla, the MR at 5K is seen to exhibit positive values. However, for the 10K MR data, a different behaviour is observed. Here, a sharp downward cusp (WL feature) is observed rather than the sharp upward cusp feature of WAL and persists up to ± 0.08Tesla exhibiting negative MR values. Above ± 0.08Tesla, the MR curve takes an upturn and further exhibits positive values above ± 0.15Tesla. Thus, a crossover from WAL to WL effect was clearly observed near zero applied magnetic fields as the temperature increased from 2K to 10K for the $Bi_{0.85}Sb_{0.15}$ sample.

Figure 10 (a, b, c) displays the magneto-conductivity (MC) analysis of $Bi_{1-x}Sb_x$ (x = 0.05, 0.1 and 0.15) samples at different temperatures using the Hikami - Larkin - Nagaoka (HLN) equation [45]:

$$\Delta\sigma(H) = \sigma(H) - \sigma(0) = -\frac{\propto e^2}{\pi h}\left[\ln\left(\frac{B_\varphi}{H}\right) - \Psi\left(\frac{1}{2} + \frac{B_\varphi}{H}\right)\right]$$

Here, $\Delta\sigma(H)$ represents the change of magneto-conductivity, α is a coefficient signifying the type of localization, e denotes the electronic charge, h represents the Planck's constant, Ψ is the digamma function, H is the applied magnetic field, $B_\varphi = \frac{h}{8e\pi H l_\varphi}$ is the characteristic magnetic field and $l_\varphi$ is the phase coherence length. As reported, for single surface conducting channel α should exhibit a value of -0.5, whereas for multi parallel conducting channels α value lies between -0.5 to -1.5 [37, 46-49]. However, experimental values of α is reported to lie between –0.4 and –1.1, for single surface state, two surface states, or intermixing between the surface and bulk states [37, 46-49]. In the present paper, we first analyzed the MC data of $Bi_{0.95}Sb_{0.05}$ sample by fitting it to the HLN equation with fields up to ± 1.5Tesla and at different temperatures viz. 2,5,10, 50, 100 and 200K [Fig. 10(a)]. The MC curve is well fitted to the HLN equation and the values of the extracted fitting parameters (α and$l_\varphi$) are displayed in Table 5. As the temperatures increases from 2 to 50K, the values of α varies from -1.145 to -1.448, indicating the existence of two decoupled surface states along with a small contribution from the bulk states, which is partially coupled with the surface states i.e., weak anti-localization (WAL) dominates the MC curve with negligible WL component. Conversely, for higher temperatures viz., 100 and 200K, the HLN analysis shows the value of α as -0.440 and -0.324 respectively, which considerably differs from -0.5, signifying the presence of a single surface state along with a major bulk contribution. The resultant α value is observed to decrease from -1.145 to -0.324, with increasing temperatures (from 2 to 200K). Accordingly, we can say that the temperature dependence of α roughly follows the MR behavior. Additionally, $l_\varphi$ is observed to decrease from 62.380nm to 35.973nm [Table 5].

We then applied the HLN formula to the MC curves of $Bi_{0.9}Sb_{0.1}$ sample measured at 2 and 5K with fields up to ± 6Tesla [Fig.10 (b)]. Accordingly, the $Bi_{0.9}Sb_{0.1}$ sample shows that for lower temperatures (2 and 5K) α varies from -1.409 to -1.498, indicating that WAL dominates the MC curve with negligible WL component. The coherence length ($l_\varphi$ ) obtained is observed to decrease from 28.762nm to 27.187nm with increase in temperature from 2 to 5K. We also performed the HLN fitting for higher temperature (10, 50 and 100K, which is not shown) and the value of α is seen to vary from -1.433 to $-4.24*10^{-5}$ suggesting much stronger tendency towards WL i.e., WL contribution is maximized and the WAL contribution is minimized. However, the HLN fit at temperatures above 5 K are not satisfactory at lower field regime. Therefore, the coherence length derived from these fits is not reliable and is not shown. The corresponding values of α and $l_\varphi$ for lower temperatures (2 and 5K) are given in Table 5.

Further, we tried to fit the MC data of $Bi_{0.85}Sb_{0.15}$ sample using the HLN formula with fields up to ± 6Tesla and at different temperatures (2, 5, 10 and 50K) which resulted in an erroneous fit. The reason is because unlike the other two samples ($Bi_{0.95}Sb_{0.05}$ and $Bi_{0.9}Sb_{0.1}$), the resultant crystal ($Bi_{0.85}Sb_{0.15}$) exhibits a different MR behaviour which is already discussed in the above paragraphs [MR section]. As reported, bulk effects start playing an important role in the MC which leads to the modified HLN equation in the high field region [50-52]. Fig. 10 (c) shows the MC curve of $Bi_{0.85}Sb_{0.15}$ sample which is fitted using the modified HLN equation:

$$\Delta\sigma(H) = \sigma(H) - \sigma(0) = -\frac{\propto e^2}{\pi h}\left[\ln\left(\frac{B_\varphi}{H}\right) - \Psi\left(\frac{1}{2} + \frac{B_\varphi}{H}\right)\right] + \beta H^2$$

where, β represents the coefficient of the quadratic term. Also, β shows the classical contribution to the MC and comprises of the contributions from characteristic magnetic fields corresponding to spin orbit scattering length ($B_{so}$) and mean free path ($B_e$) in the high field region [50-52]. The fitting was done at 2, 5 and 10K for the entire field range (up to ±6Tesla). The MC data for 2, 5 and 10K fits well to the modified HLN equation and the corresponding values of α, $l_\varphi$, β and $R^2$ are displayed in Table 5. However, the MC data for 50K is not shown as it resulted in an erroneous fit. From table 5, it can be seen that the value of α varies from -0.011 to -0.003; whereas the phase coherence length varies from 17.330 to 23.083nm as the temperature increases from 2 to 10K.

## Conclusion

Summarily, a series of $Bi_{1-x}Sb_x$ single crystals ranging from x = 0.05 to x = 0.15 was synthesized using the facile self flux method, and its physical and magneto-transport properties were studied. XRD pattern revealed the growth of synthesized $Bi_{1-x}Sb_x$ sample along (00l) plane, whereas the SEM along with EDAX measurements displayed the layered structure with near stoichiometric composition, without foreign contamination. The Raman scattering studies displayed a total of four $E_g$ and $A_{1g}$ peaks corresponding to Bi-Bi and Sb-Sb vibrations vibrational modes for the studied $Bi_{1-x}Sb_x$ system. The temperature dependent electrical resistivity measurements illustrated the metallic nature of the as synthesized $Bi_{1-x}Sb_x$ system. Furthermore, as grown $Bi_{1-x}Sb_x$ samples with nominal x values of 0.05 and 0.1 ($Bi_{0.95}Sb_{0.05}$ and $Bi_{0.9}Sb_{0.1}$) exhibited a linear positive MR value whereas, the other $Bi_{1-x}Sb_x$ sample ($Bi_{0.85}Sb_{0.15}$) displays an asymptotic type of behaviour at higher applied magnetic fields (say above 5Tesla). Clearly, as the alloy composition increases from 0.05 to 0.15 the MR (%) value is observed to decrease from 1400% for $Bi_{0.95}Sb_{0.05}$ to 440% for $Bi_{0.9}Sb_{0.1}$ and then finally to 57% for $Bi_{0.85}Sb_{0.15}$ at 2K respectively. Also, as the temperature and doping concentration increases, the value of $\alpha$ and $l_\varphi$ is seen to decrease. Our result indicates the coexistence of negative MR and WAL/WL behaviour at lower applied magnetic field (below ± 0.2Tesla), which seems interesting and can open a new field of potential applications in the existing TI system.

**Acknowledgements**

The authors from CSIR-NPL would like to thank their Director NPL, India, for his keen interest in the present work. Authors further thank Mrs. Shaveta Sharma for Raman studies. S. Patnaik thanks DST-SERB project (EMR/2016/003998) for the low temperature high magnetic facility at JNU, New Delhi. Rabia Sultana and Ganesh Gurjar thank CSIR, India, for research fellowship. Rabia Sultana thanks AcSIR-NPL for Ph.D. registration.

**Figure Captions**

**Figure 1:** Schematic heat treatment diagram for $Bi_{1-x}Sb_x$ single crystal.

**Figure 2:** Image of as grown and mechanically cleaved $Bi_{1-x}Sb_x$ single crystal.

**Figure 3:** X-ray diffraction pattern of as synthesized (a) $Bi_{0.95}Sb_{0.05}$ (b) $Bi_{0.9}Sb_{0.1}$ and (c) $Bi_{0.85}Sb_{0.15}$ single crystals.

**Figure 4:** Rietveld fitted room temperature XRD pattern for powder (a) $Bi_{0.95}Sb_{0.05}$ (b) $Bi_{0.9}Sb_{0.1}$ and (c) $Bi_{0.85}Sb_{0.15}$ samples.

**Figure 5:** Unit cell structure of (a) $Bi_{0.95}Sb_{0.05}$ (b) $Bi_{0.9}Sb_{0.1}$ and (c) $Bi_{0.85}Sb_{0.15}$ single crystals.

**Figure 6:** SEM images and EDXA mapping of (a) $Bi_{0.95}Sb_{0.05}$ (b) $Bi_{0.9}Sb_{0.1}$ and (c) $Bi_{0.85}Sb_{0.15}$ single crystals.

**Figure 7:** Room temperature Raman spectra of (a) $Bi_{0.95}Sb_{0.05}$ (b) $Bi_{0.9}Sb_{0.1}$ and (c) $Bi_{0.85}Sb_{0.15}$ single crystals.

**Figure 8:** Temperature dependent electrical resistivity under different applied magnetic fields for (a) $Bi_{0.95}Sb_{0.05}$ (b) $Bi_{0.9}Sb_{0.1}$ and (c) $Bi_{0.85}Sb_{0.15}$ single crystal in a temperature range of 300K to 5K. Inset (a, b, c) shows the Fermi Liquid behaviour of (a) $Bi_{0.95}Sb_{0.05}$ (b) $Bi_{0.9}Sb_{0.1}$ and (c) $Bi_{0.85}Sb_{0.15}$ single crystals at 0Tesla.

**Figure 9:** MR (%) as a function of magnetic field (H) at different temperatures for (a) $Bi_{0.95}Sb_{0.05}$ (b) (i) $Bi_{0.9}Sb_{0.1}$ (ii) Zoom in view of $Bi_{0.9}Sb_{0.1}$ (c) (i) $Bi_{0.85}Sb_{0.15}$ (ii) Quadratic fitting of $Bi_{0.85}Sb_{0.15}$ (iii) Zoom in view of $Bi_{0.85}Sb_{0.15}$ single crystals.

**Figure 10:** WAL related magneto-conductivity for (a) $Bi_{0.95}Sb_{0.05}$ and (b) $Bi_{0.9}Sb_{0.1}$ single crystal at various temperatures fitted using the HLN equation. (c) MC for $Bi_{0.85}Sb_{0.15}$ single crystal at various temperatures fitted using the modified HLN equation

## Table Captions

**Table 1:** Rietveld refined lattice parameters, atom parameters as well as the Chi square values of $Bi_{1-x}Sb_x$ (x=0.05, 0.1 and 0.15) system.

**Table 2:** Vibrational mode values of $Bi_{1-x}Sb_x$ (x = 0.05, 0.1 and 0.15) system.

**Table 3:** Residual resistivity ($\rho_0$) and constant (A) values obtained from the Fermi liquid fitting for $Bi_{1-x}Sb_x$ (x = 0.05, 0.1 and 0.15) system.

**Table 4:** MR (%) values at different temperatures under 6Tesla applied magnetic field for $Bi_{1-x}Sb_x$ (x = 0.05, 0.1 and 0.15) system.

**Table 5:** HLN fit values of the pre-factor ($\alpha$), phase coherence length ($l\phi$), quadratic term ($\beta$) and $R^2$ at different temperatures for $Bi_{1-x}Sb_x$ (x = 0.05, 0.1 and 0.15) system.

Table 1

| Alloy Composition | a (Å) | b (Å) | c (Å) | $\chi^2$ | Atom Parameters | | |
|---|---|---|---|---|---|---|---|
| | | | | | x | y | z |
| $Bi_{0.95}Sb_{0.05}$ | 4.521(7) | 4.523(7) | 11.81(2) | 5.76 | 0 | 0 | 0.233 |
| $Bi_{0.9}Sb_{0.1}$ | 4.512(4) | 4.512(4) | 11.77(8) | 2.96 | 0 | 0 | 0.234 |
| $Bi_{0.85}Sb_{0.15}$ | 4.514(5) | 4.514(5) | 11.77(5) | 6.35 | 0 | 0 | 0.233 |

Table 2

| Alloy Composition | Vibrational modes of $Bi_{1-x}Sb_x$ system | | | |
|---|---|---|---|---|
| | Bi-Bi vibrations | | Sb-Sb vibrations | |
| | $E_g$ ($cm^{-1}$) | $A_{1g}$ ($cm^{-1}$) | $E_g$ ($cm^{-1}$) | $A_{1g}$ ($cm^{-1}$) |
| $Bi_{0.95}Sb_{0.05}$ | 71.709 | 96.603 | 117.585 | 138.150 |
| $Bi_{0.9}Sb_{0.1}$ | 72.947 | 99.209 | 120.141 | 138.696 |
| $Bi_{0.85}Sb_{0.15}$ | 75.105 | 99.929 | 119.980 | 140.909 |

Table 3

| $\rho = \rho_0 + AT^2$ | | |
|---|---|---|
| **Alloy Composition** | $\rho_0$ | A |
| $Bi_{0.95}Sb_{0.05}$ | $5.148*10^{-5}$ | $6.706*10^{-9}$ |
| $Bi_{0.9}Sb_{0.1}$ | $4.097*10^{-5}$ | $5.405*10^{-9}$ |
| $Bi_{0.85}Sb_{0.15}$ | $4.658*10^{-5}$ | $3.129*10^{-9}$ |

Table 4

| Temperature | MR (%) | | |
|---|---|---|---|
| | $Bi_{0.95}Sb_{0.05}$ (up to ± 6Tesla) | $Bi_{0.9}Sb_{0.1}$ (up to ± 6Tesla) | $Bi_{0.85}Sb_{0.15}$ (up to ± 6Tesla) |
| 2K | ~1400 | ~440 | ~60 |
| 5K | ~1320 | ~430 | ~90 |
| 10K | ~1310 | ~415 | ~110 |
| 50K | ~1170 | ~295 | ~52 |
| 100K | ~775 | ~205 | |
| 200K | ~500 | ~130 | |

Table 5

| Temperature | $Bi_{0.95}Sb_{0.05}$ (up to ± 1.5Tesla) | | | $Bi_{0.9}Sb_{0.1}$ (up to ± 6Tesla) | | | $Bi_{0.85}Sb_{0.15}$ (up to ± 6Tesla) | | | |
|---|---|---|---|---|---|---|---|---|---|---|
| | α | $l_{\varphi}$ (nm) | $R^2$ | α | $l_{\varphi}$ (nm) | $R^2$ | α | $l_{\varphi}$ (nm) | β | $R^2$ |
| 2K | -1.145 | 62.380 | 0.980 | -1.409 | 28.762 | 0.986 | -0.011 | 17.330 | $-1.808*10^{-4}$ | 0.984 |
| 5K | -1.233 | 55.654 | 0.976 | -1.498 | 27.187 | 0.986 | -0.007 | 18.840 | $-8.299*10^{-5}$ | 0.988 |
| 10K | -1.498 | 47.044 | 0.976 | | | | -0.003 | 23.083 | $-1.3810*10^{-4}$ | 0.981 |
| 50K | -1.448 | 36.521 | 0.987 | | | | | | | |
| 100K | -0.440 | 51.578 | 0.999 | | | | | | | |
| 200K | -0.324 | 35.973 | 0.936 | | | | | | | |

Fig. 1

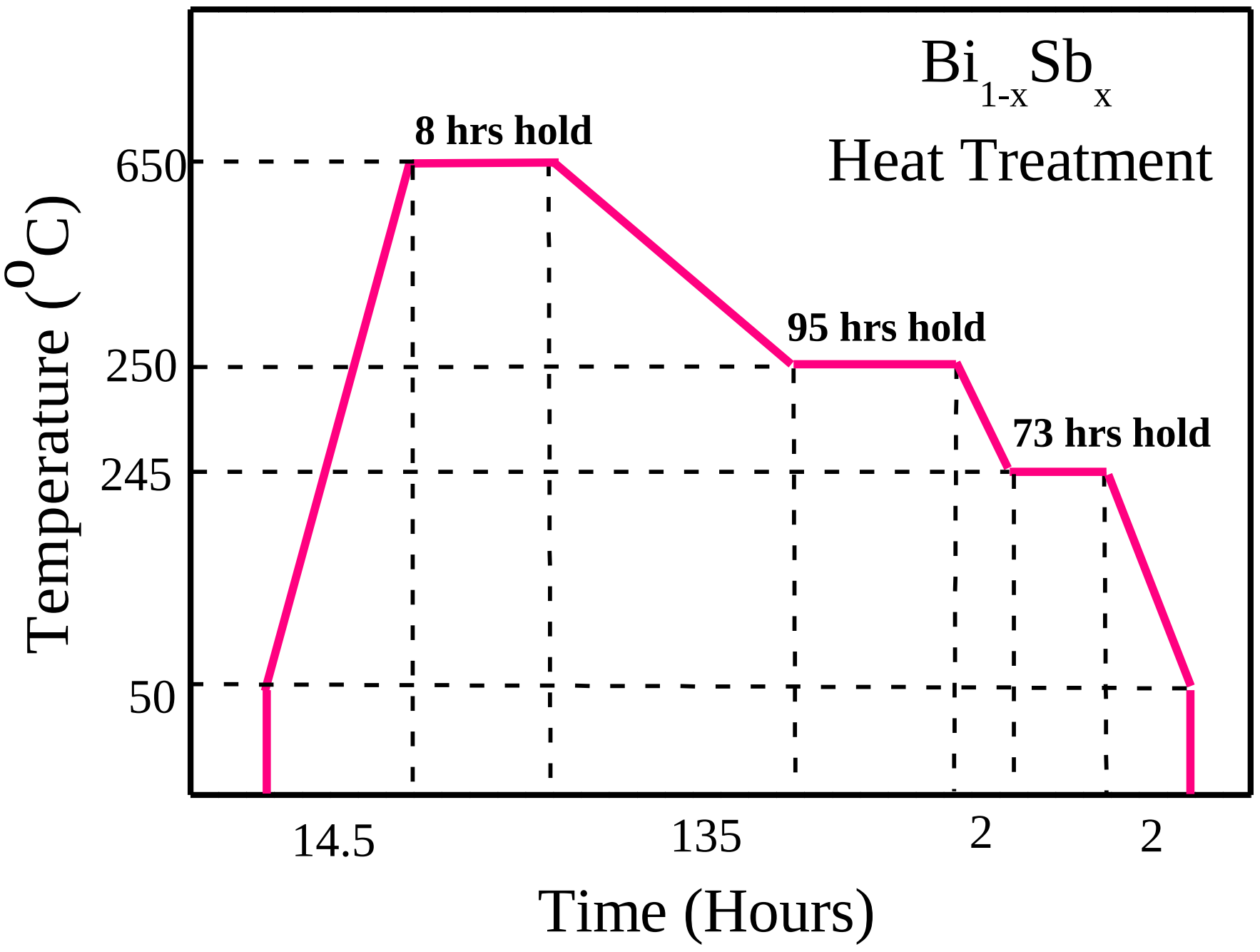

Fig. 2

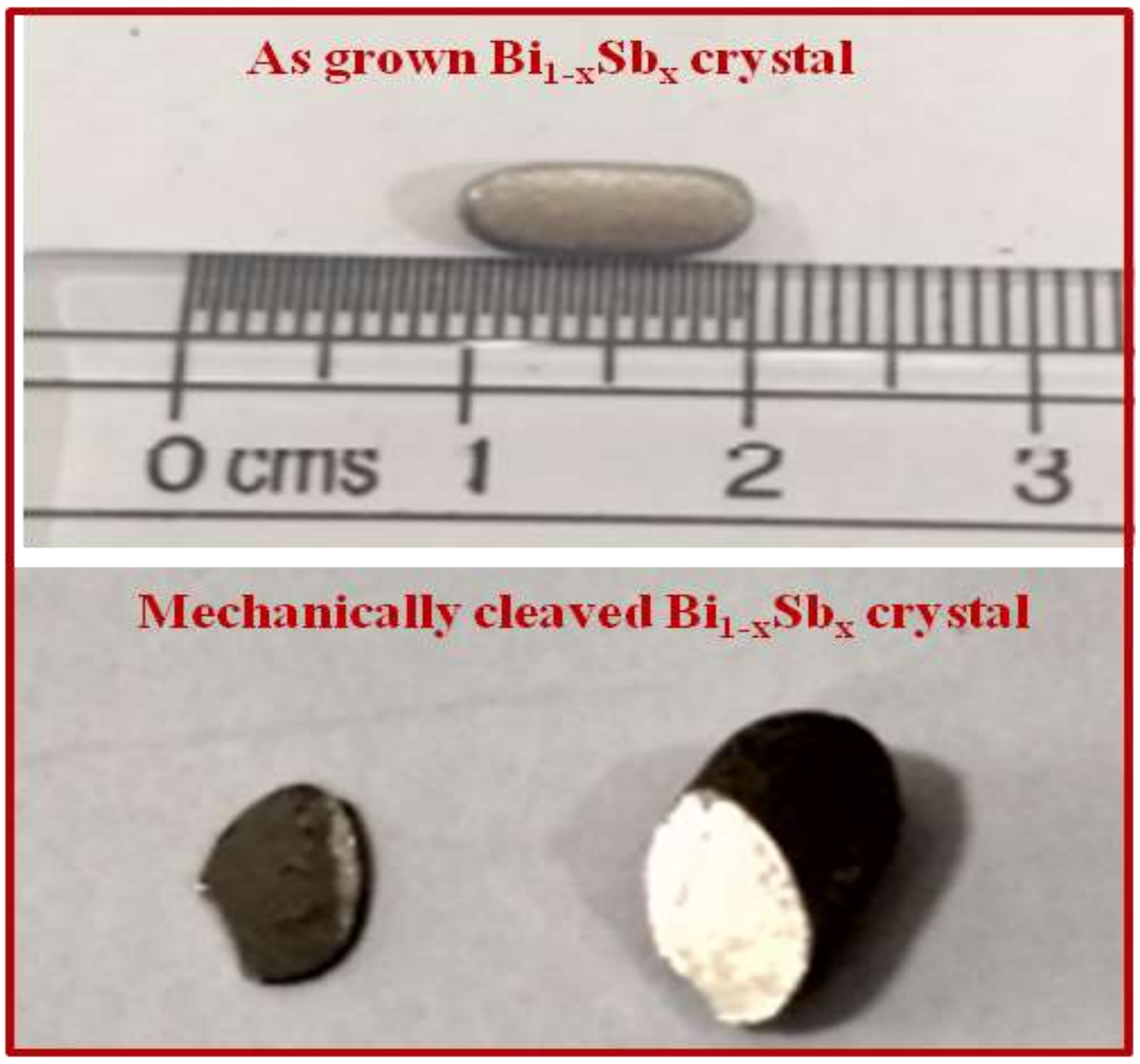

As grown Bi1-xSbx crystal
0 cms 1 2 3
Mechanically cleaved Bi1-xSbx crystal

Fig. 3

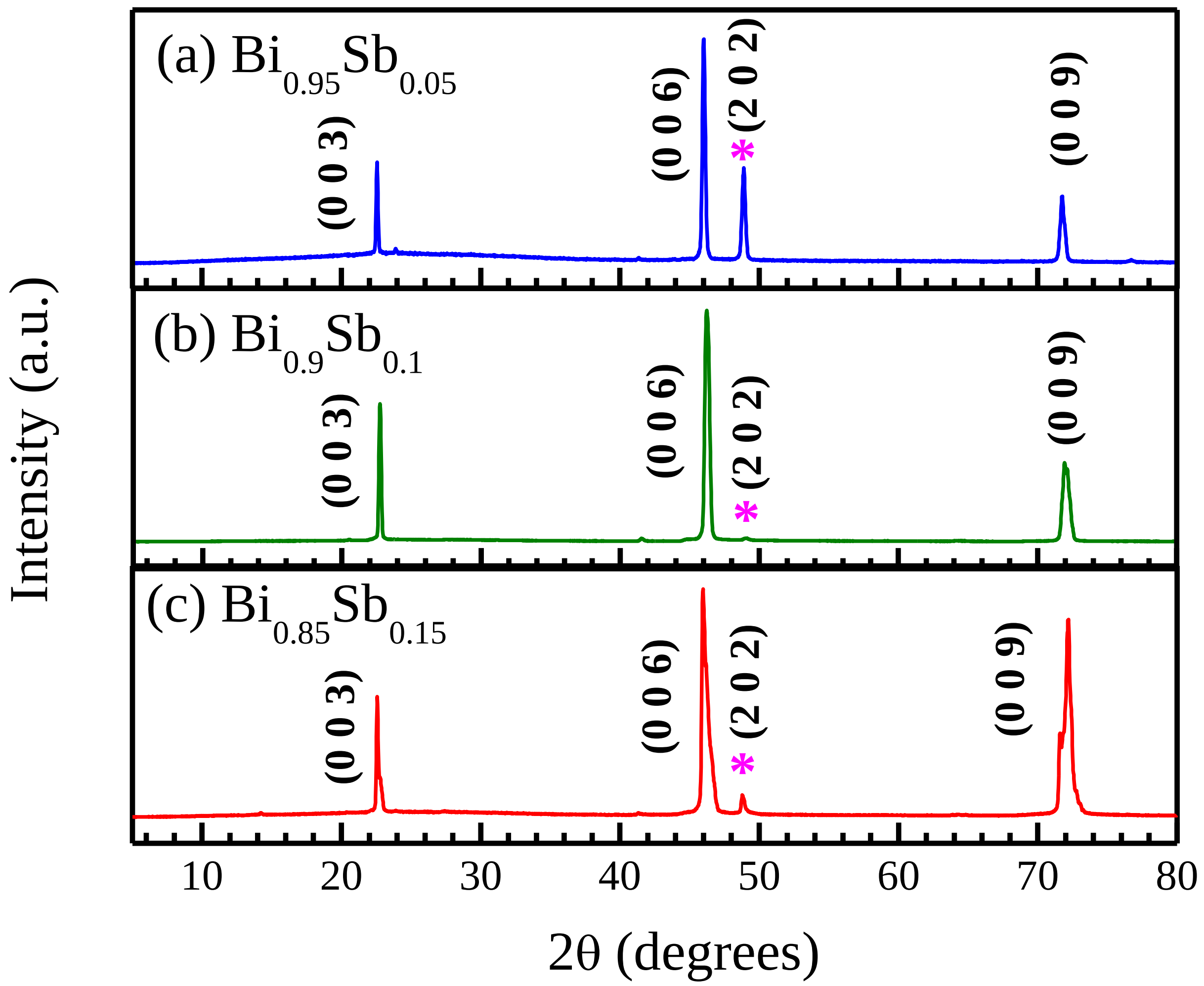

Fig. 4

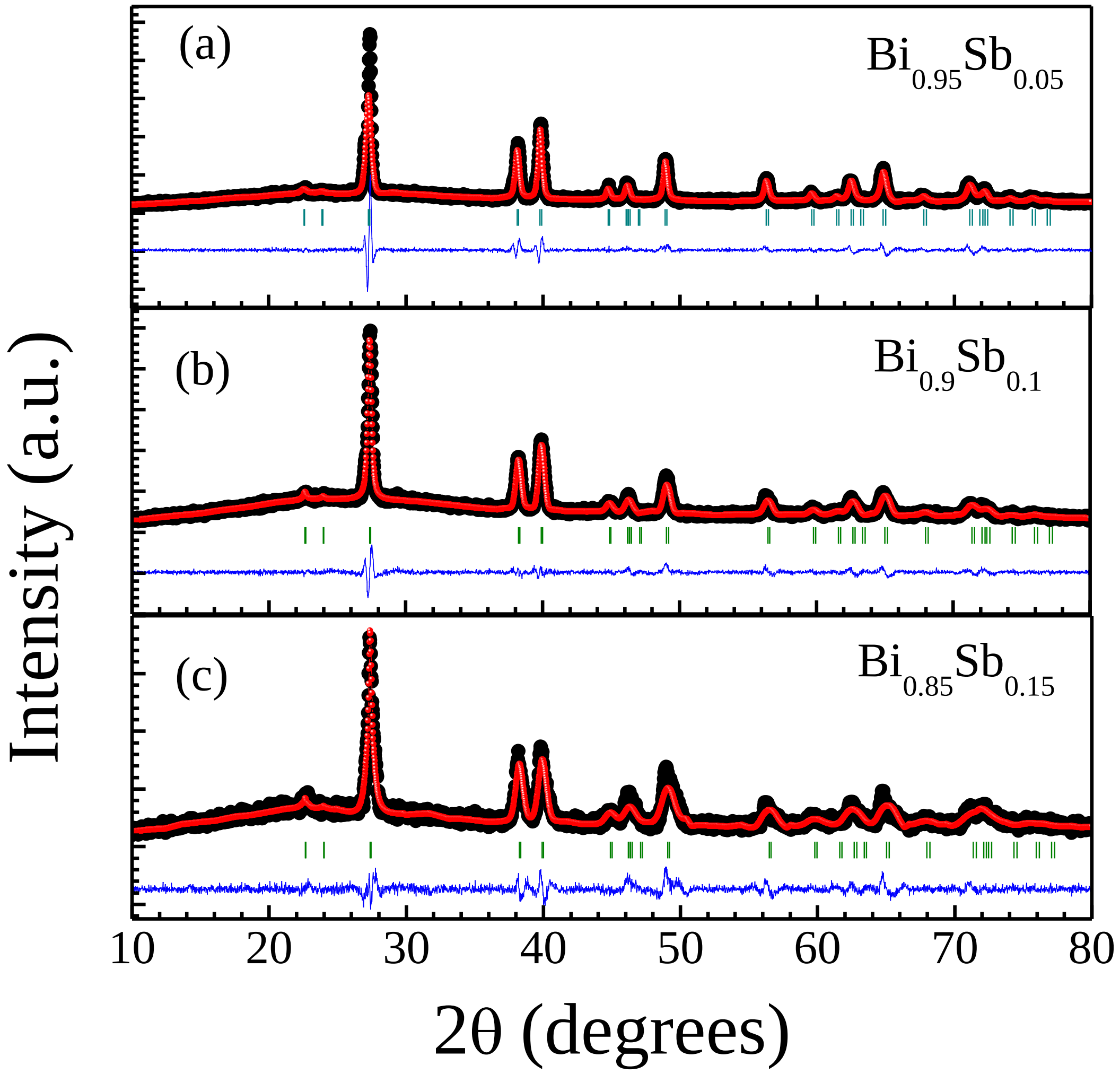

Fig. 5

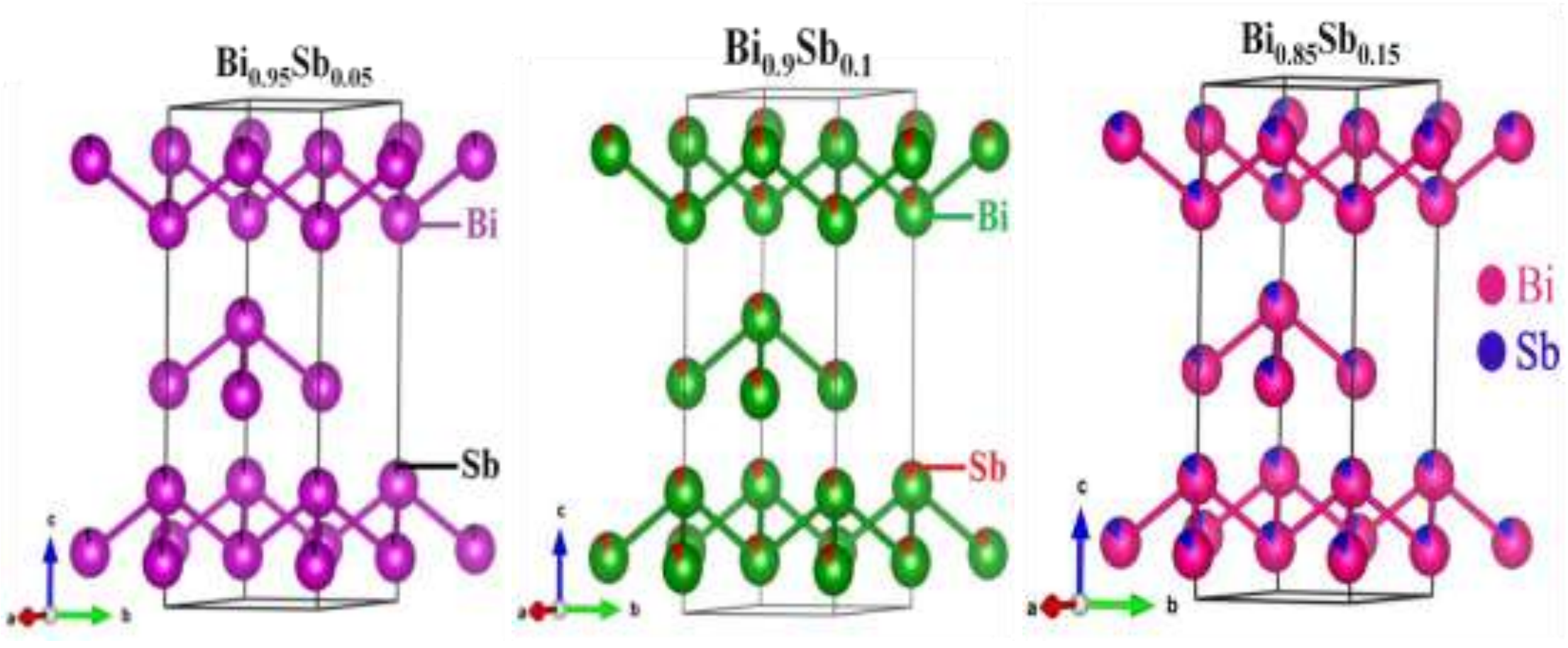

Fig 6(a)

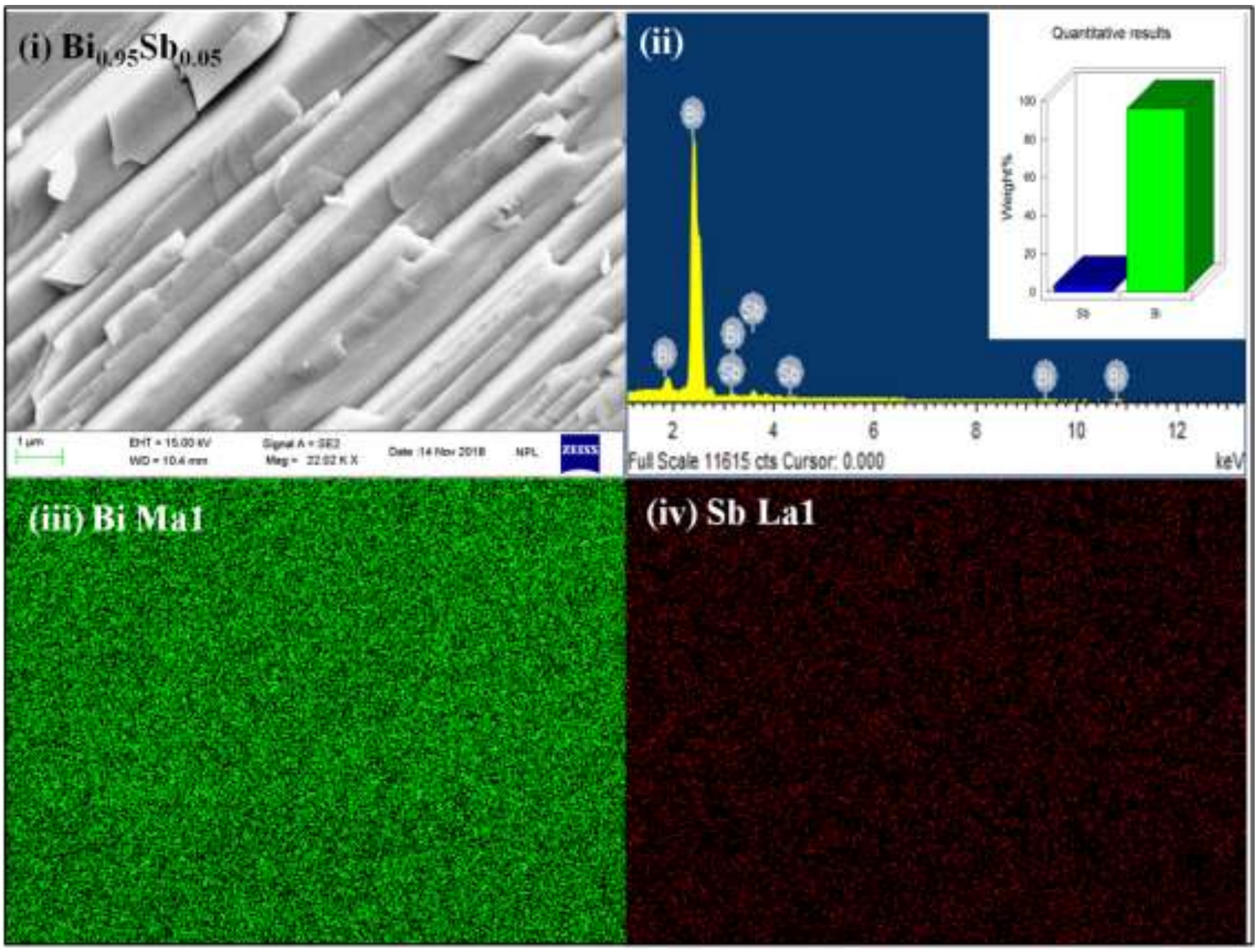

Fig 6 (b)

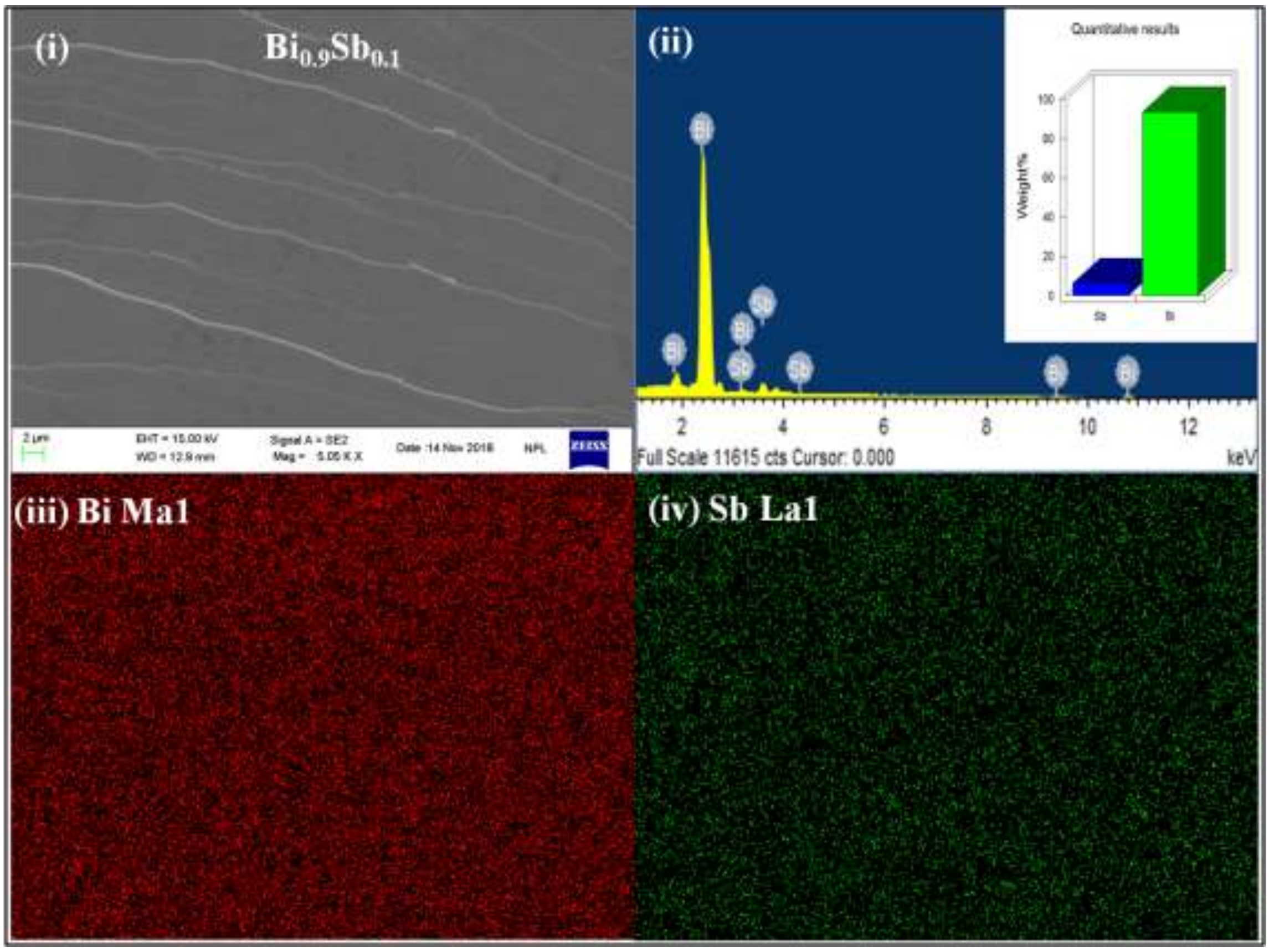

Fig.6(c)

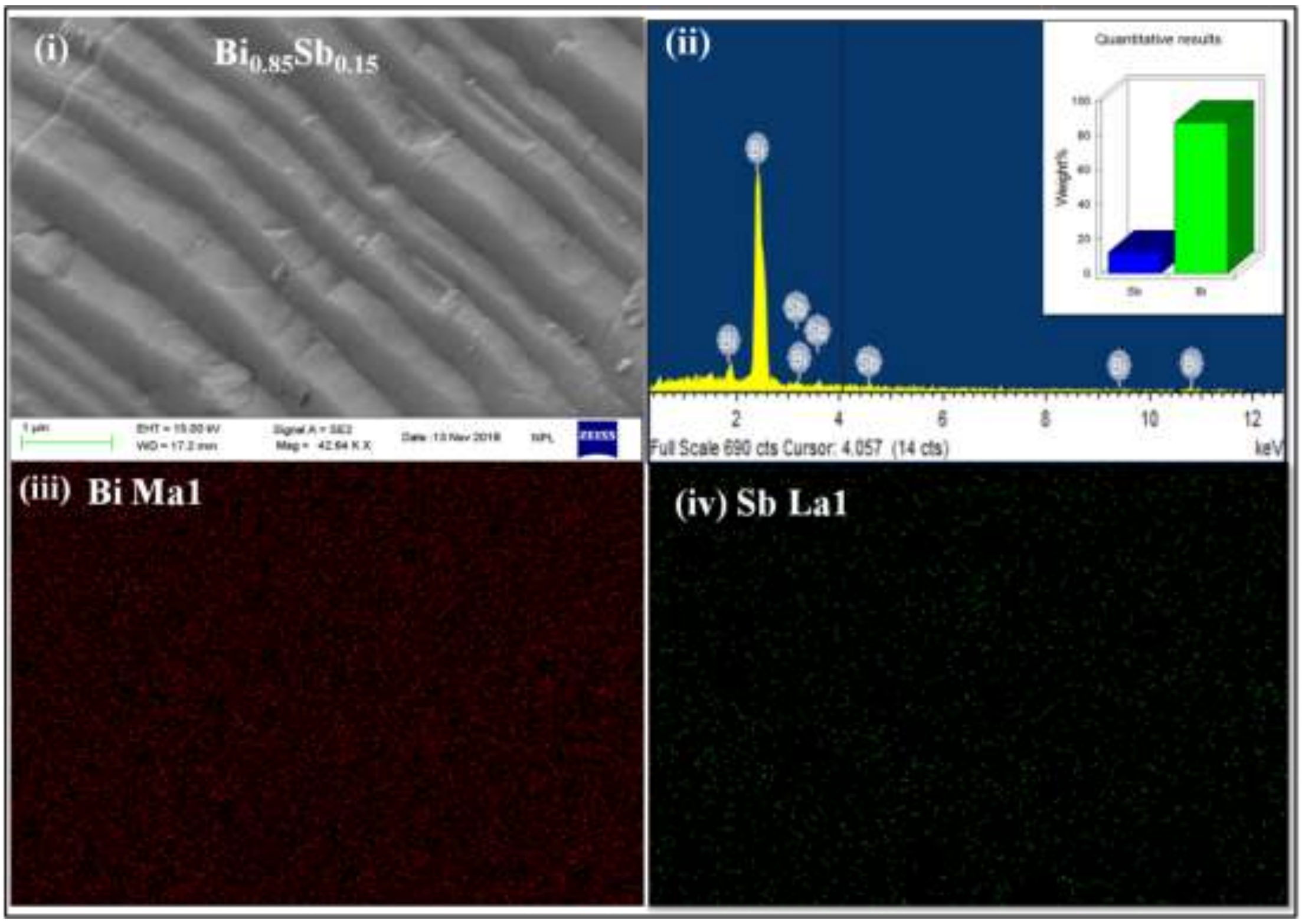
(i) $Bi_{0.85}Sb_{0.15}$
(ii)
Quantitative results
Full Scale 690 cts Cursor: 4.057 (14 cts)
keV
(iii) Bi Ma1
(iv) Sb La1

Fig. 7

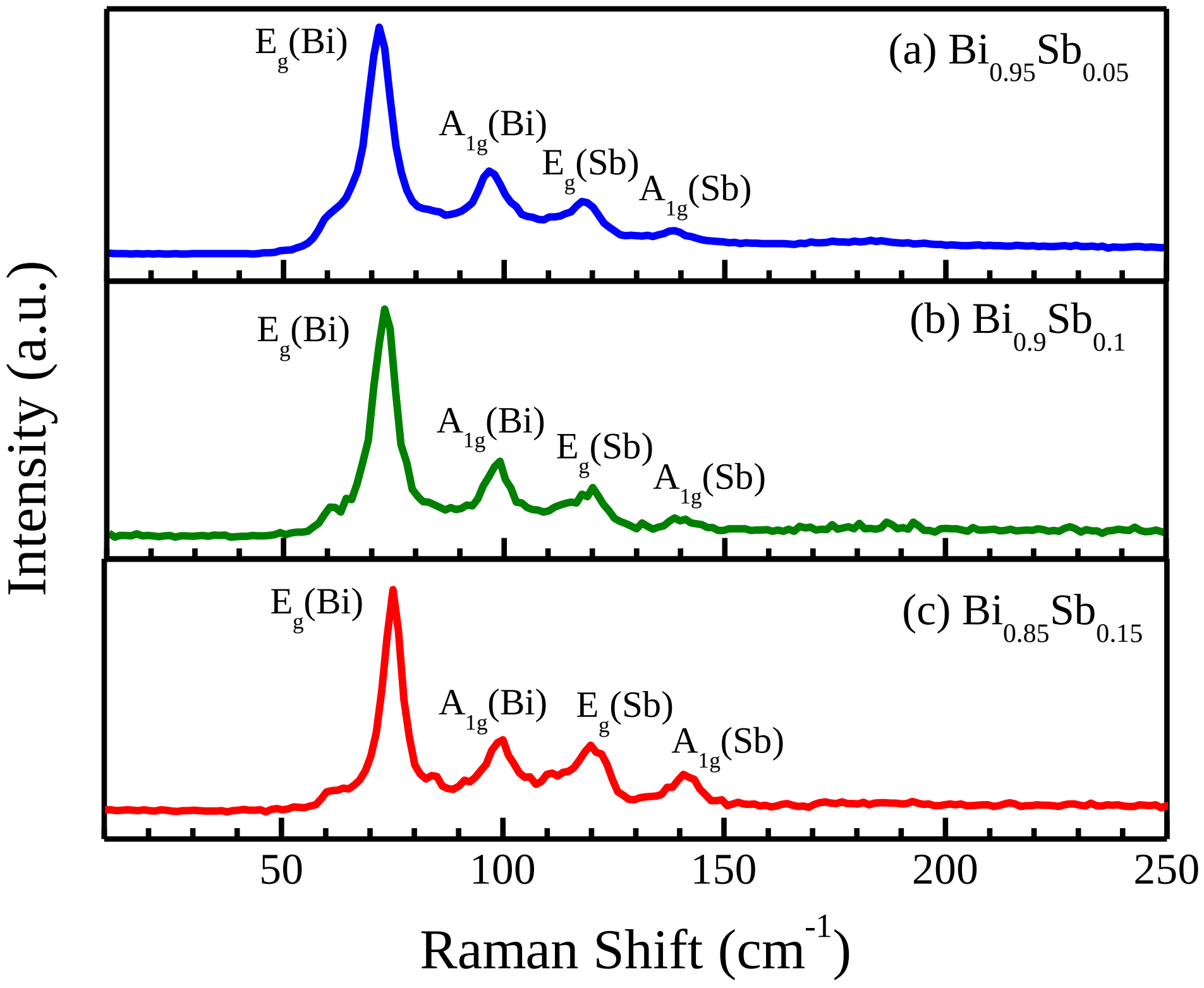

Fig. 8(a)

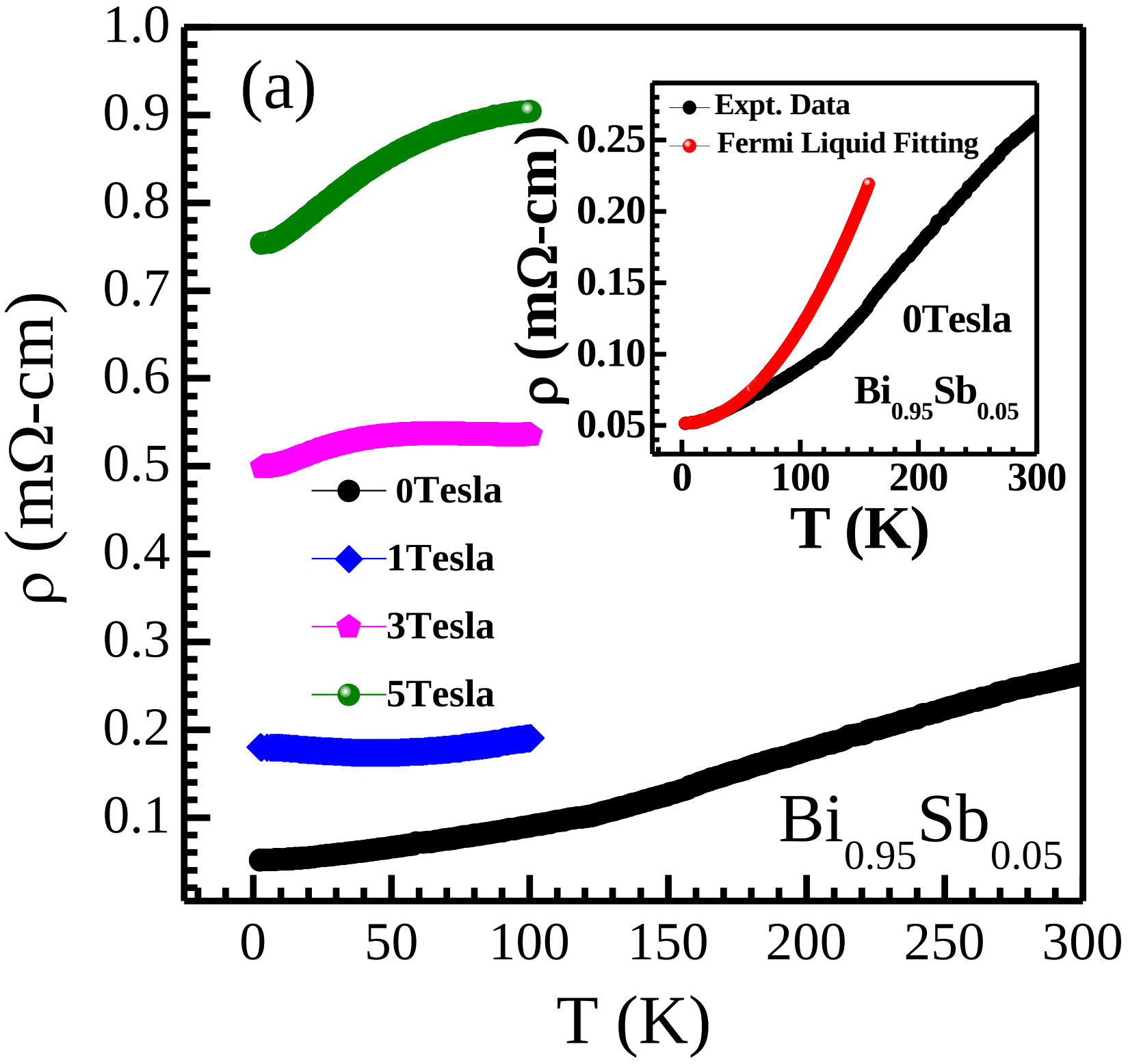

Fig. 8(b)

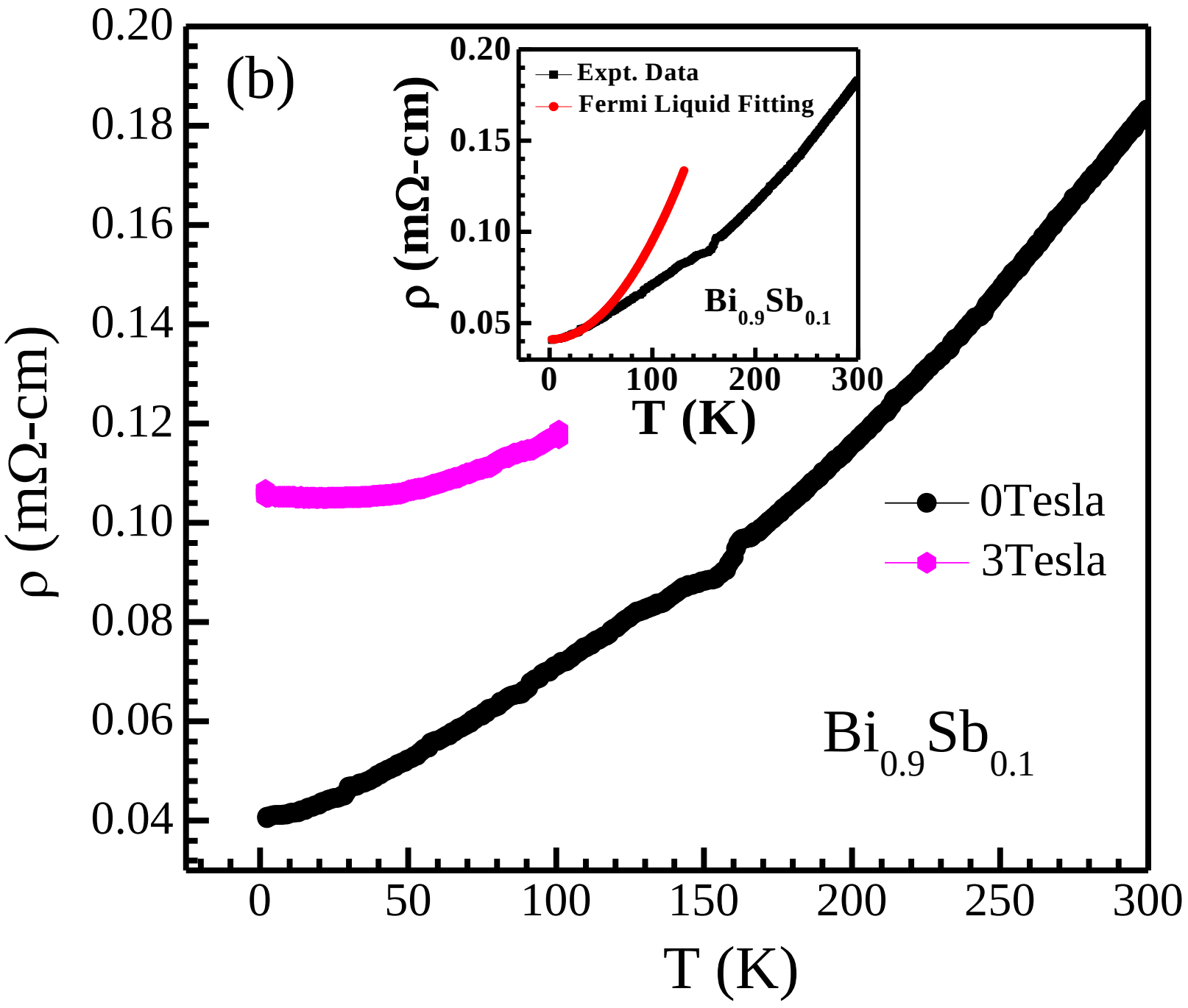

Fig. 8(c)

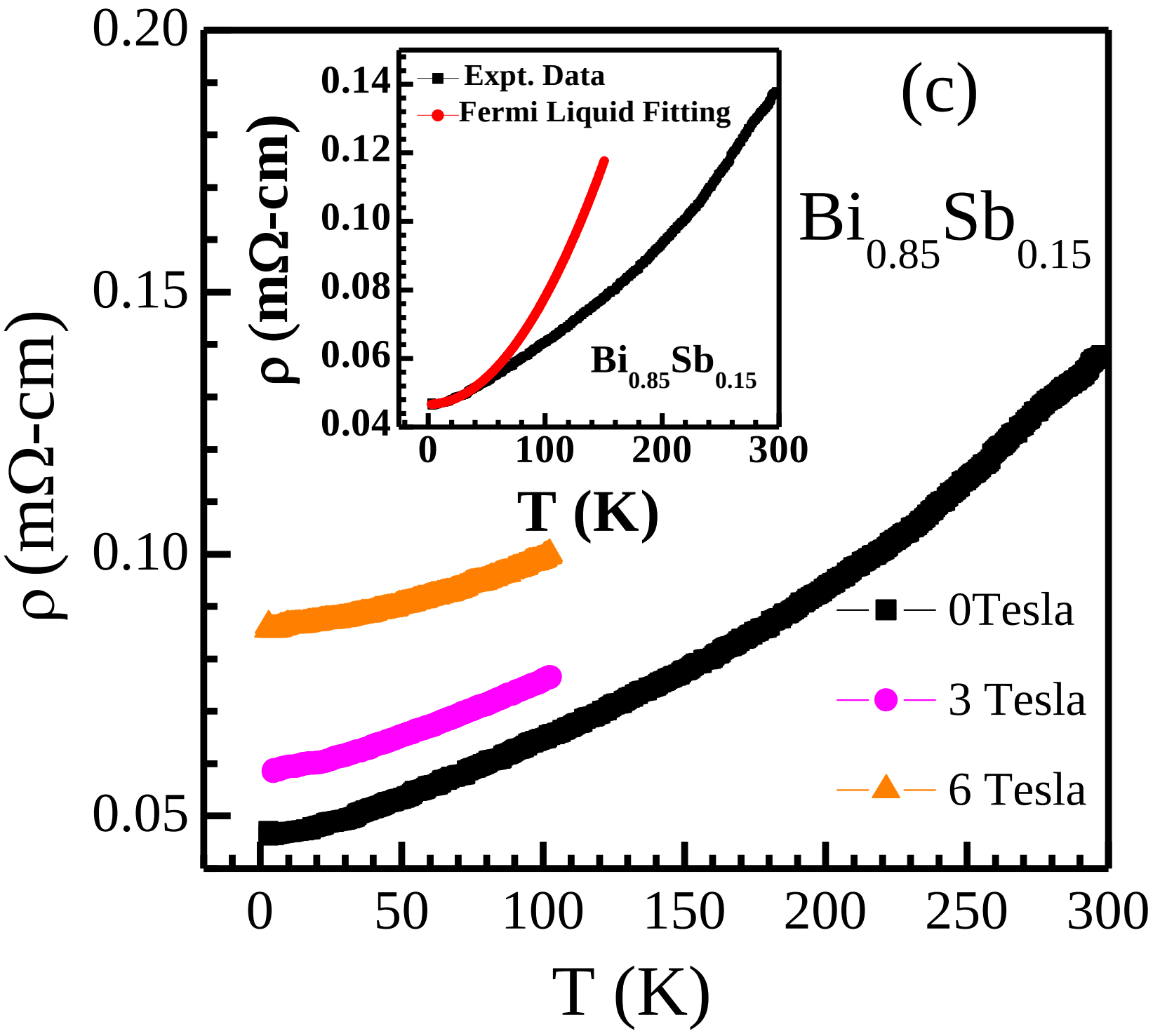

Fig. 9(a)

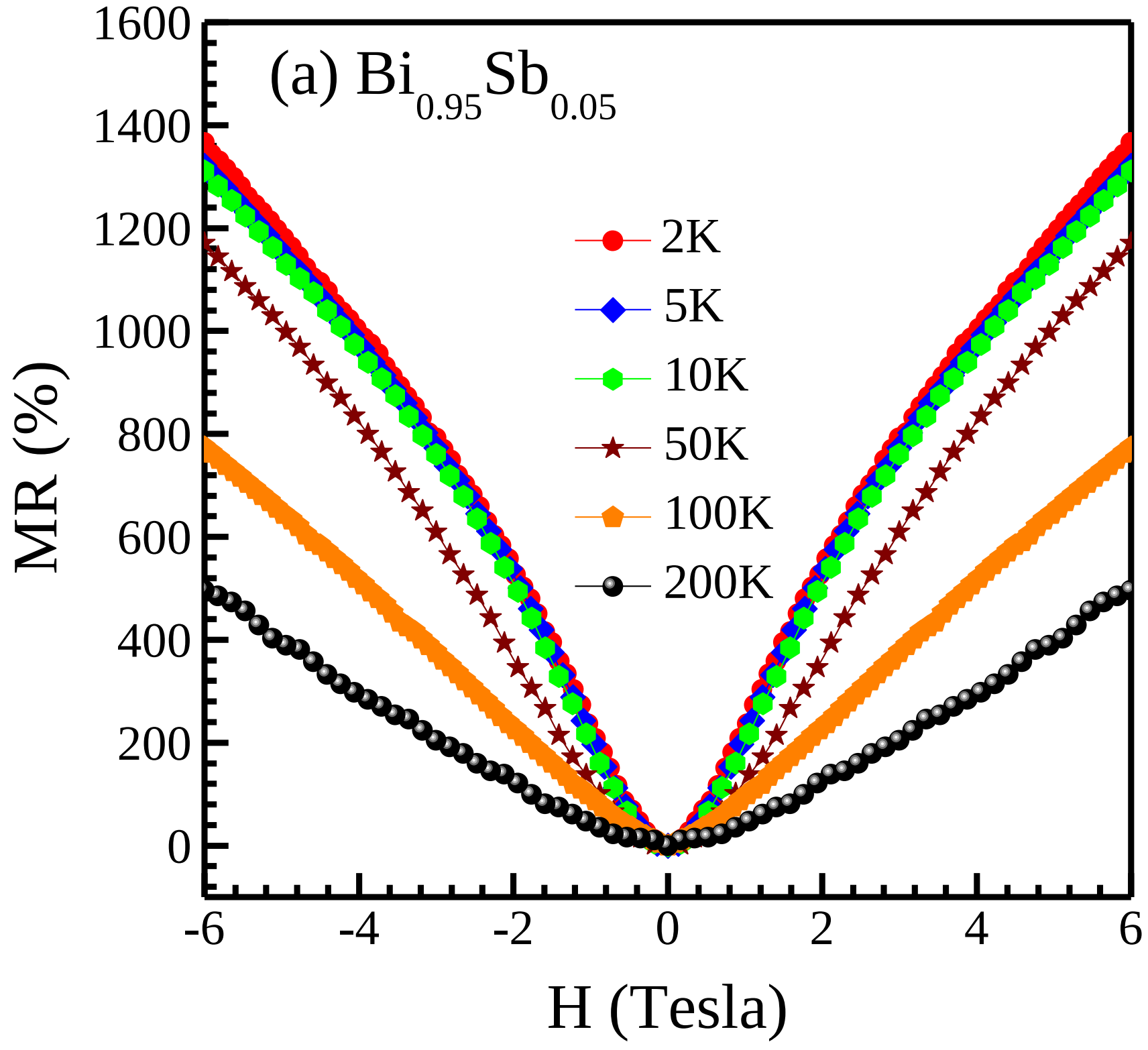

Fig. 9(b)

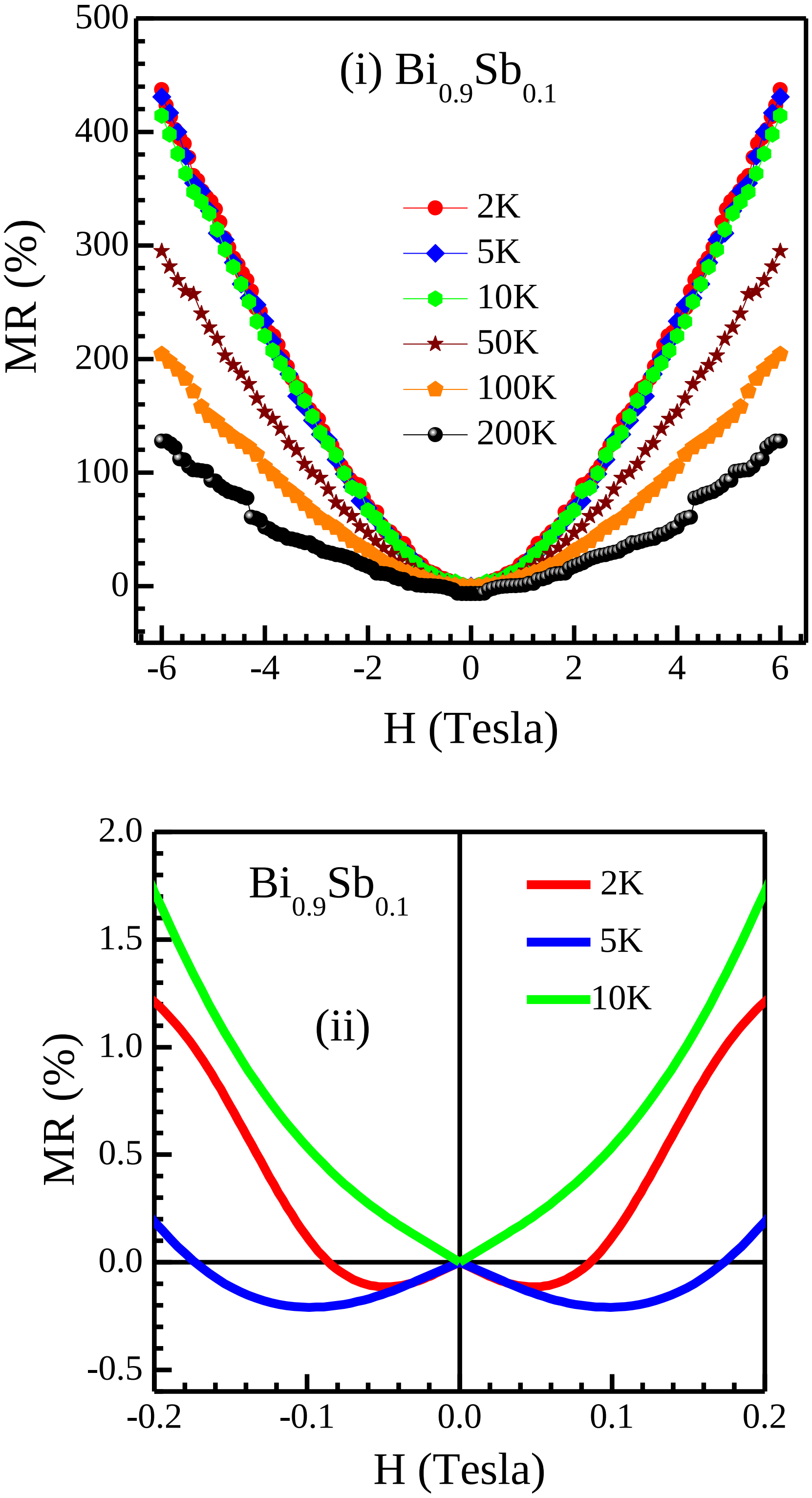

Fig. 9(c)

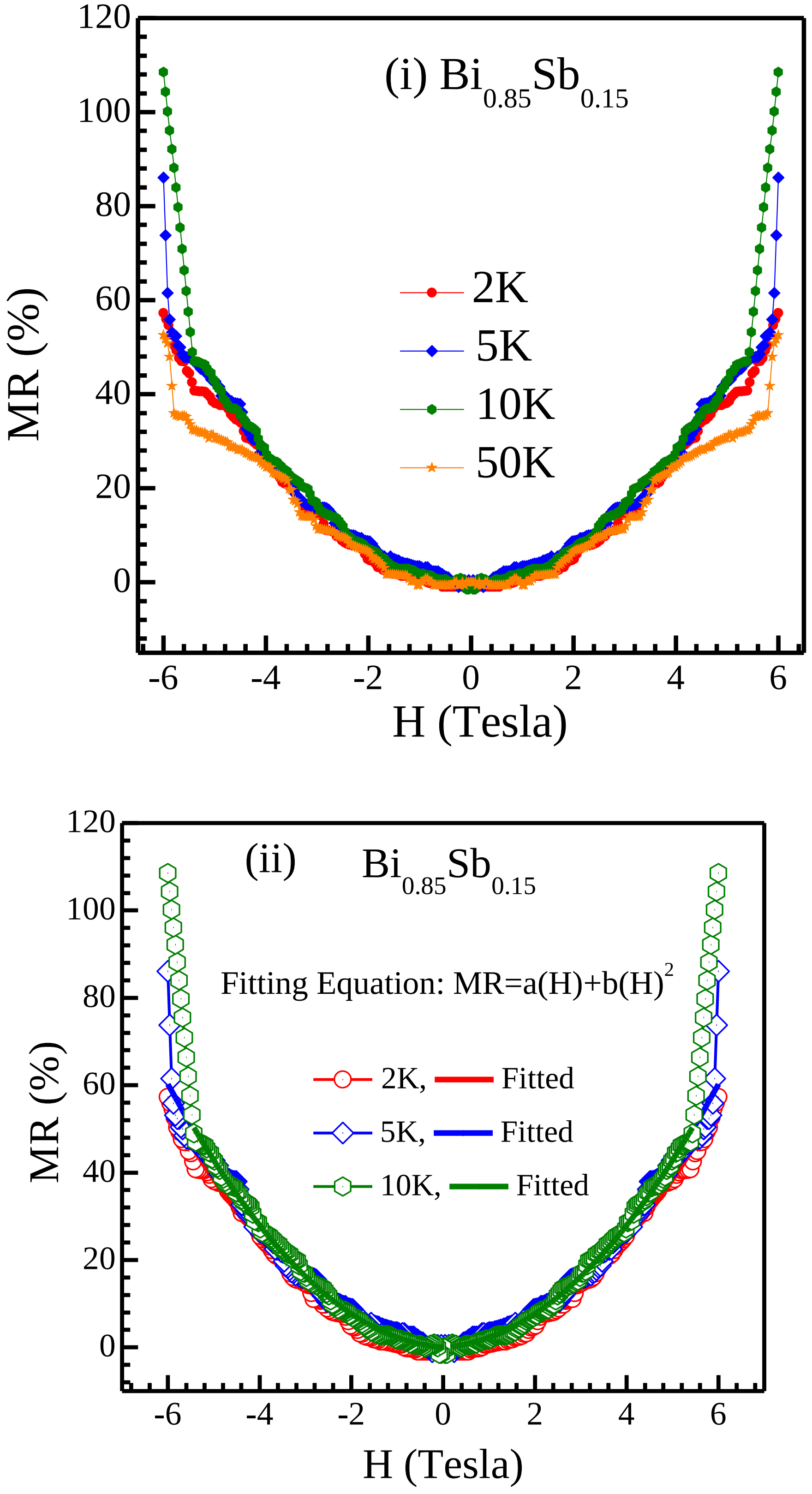

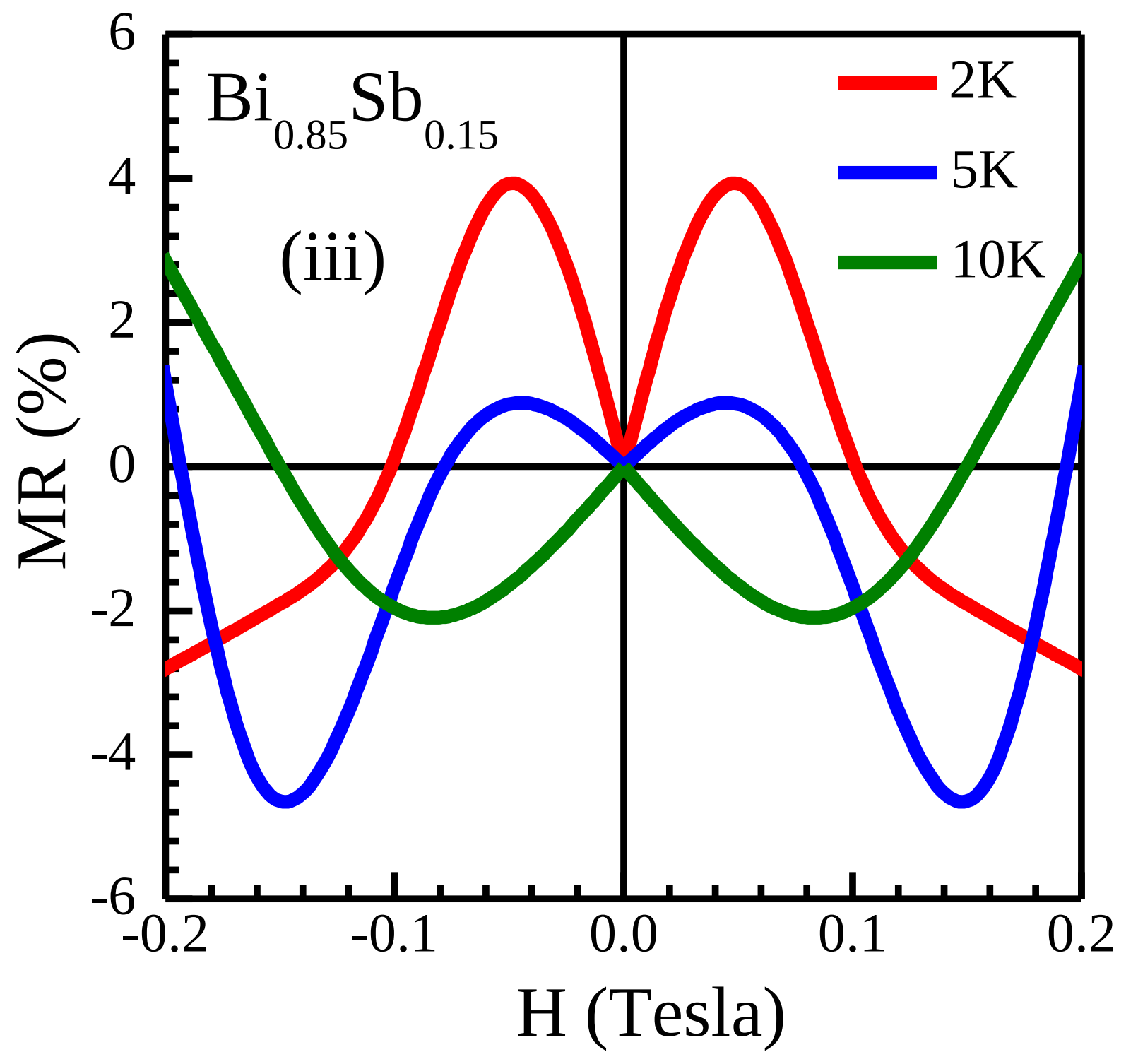

Fig. 10(a)

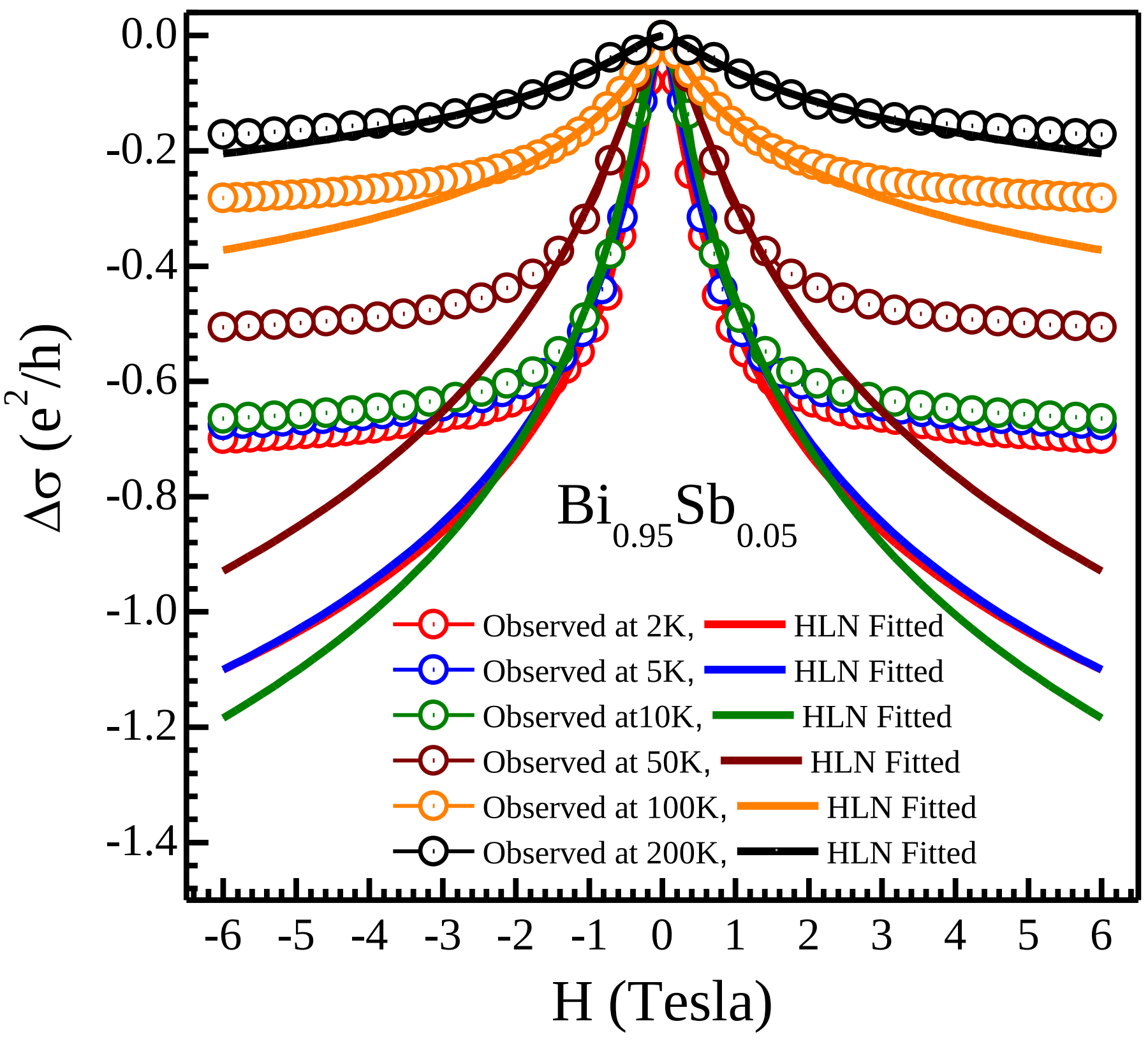

$Bi_{0.95}Sb_{0.05}$
Observed at 2K, HLN Fitted
Observed at 5K, HLN Fitted
Observed at10K, HLN Fitted
Observed at 50K, HLN Fitted
Observed at 100K, HLN Fitted
Observed at 200K, HLN Fitted
$\Delta\sigma$ ($e^2$/h)
H (Tesla)

Fig. 10(b)

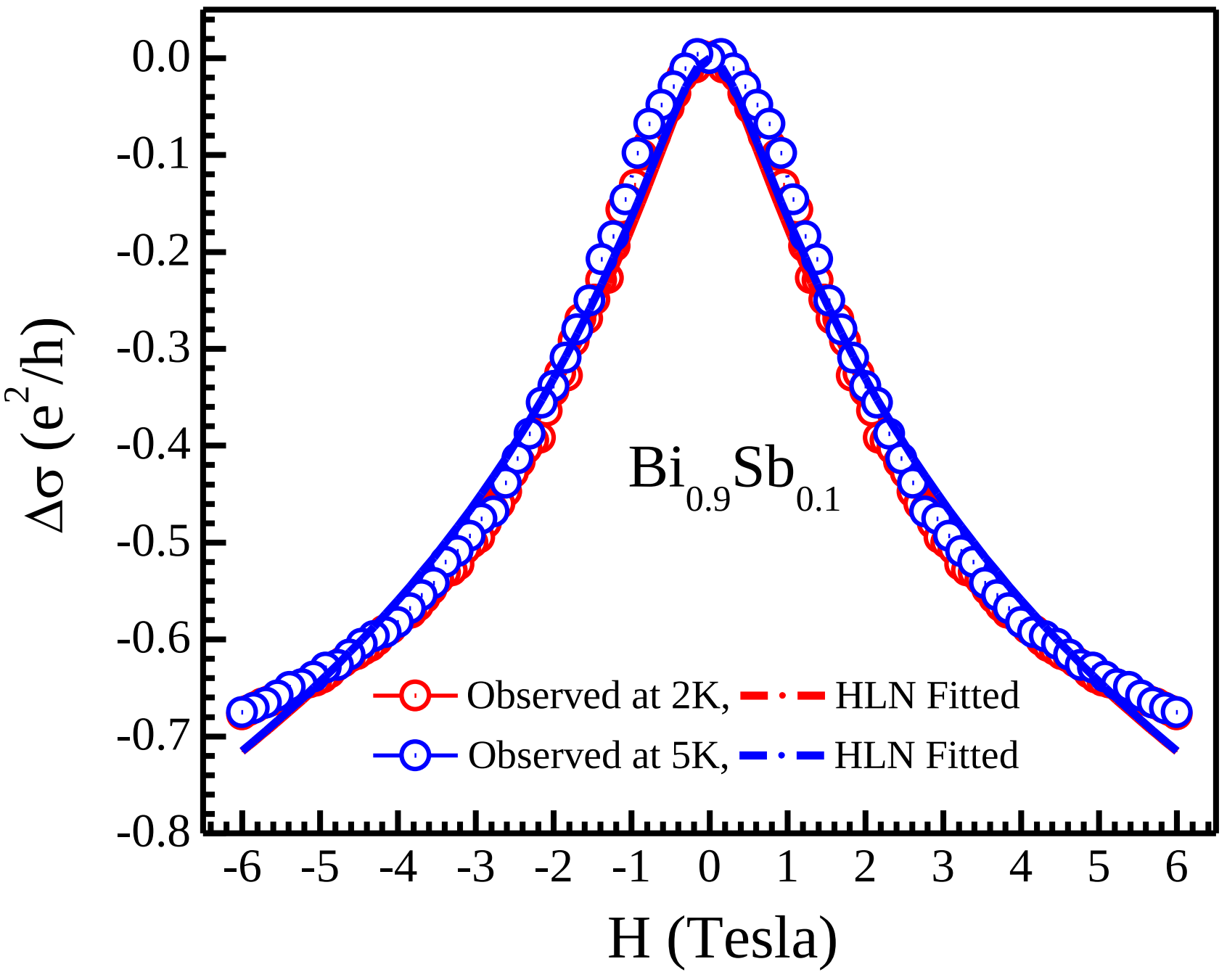

Fig. 10(c)

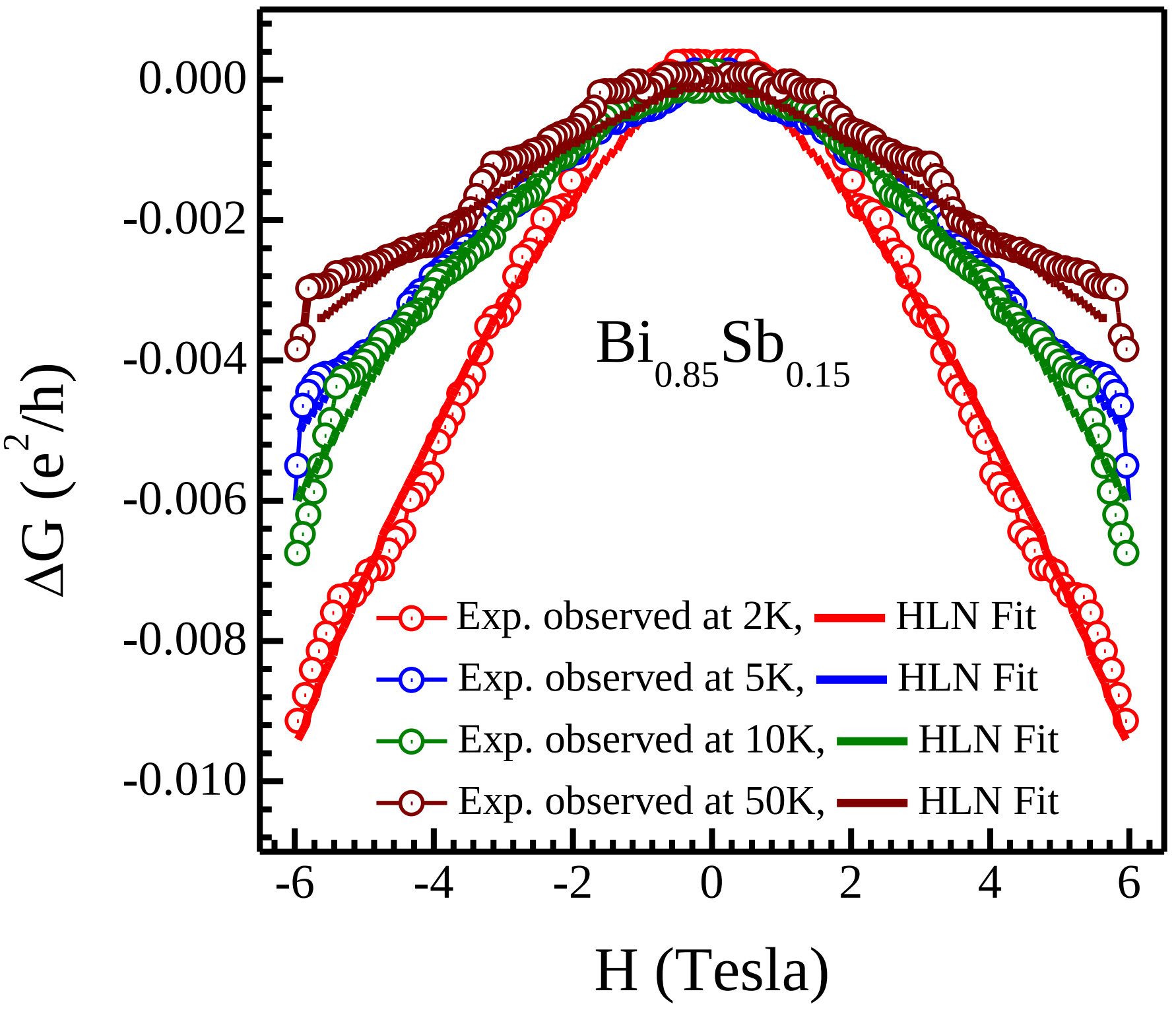